\documentclass[%
reprint,
superscriptaddress,
 amsmath,amssymb,
 aps,
prd,
]{revtex4-2}

\usepackage{graphicx}
\usepackage{dcolumn}
\usepackage{bm}

\begin{document}


\title{Alternative LISA-TAIJI networks: detectability of the isotropic stochastic gravitational wave background}


\author{Gang Wang}
\email[Gang Wang: ]{gwang@shao.ac.cn, gwanggw@gmail.com}
\affiliation{Shanghai Astronomical Observatory, Chinese Academy of Sciences, Shanghai 200030, China}

\author{Wen-Biao Han}
\affiliation{Shanghai Astronomical Observatory, Chinese Academy of Sciences, Shanghai 200030, China}
\affiliation{Hangzhou Institute for Advanced Study, University of Chinese Academy of Sciences, Hangzhou 310124, China}
\affiliation{School of Astronomy and Space Science, University of Chinese Academy of Sciences, Beijing 100049, China}

\date{\today}

\begin{abstract}

In previous work \cite{Wang:alternative}, three TAIJI orbital deployments have been proposed to compose alternative LISA-TAIJI networks, TAIJIm (leading the Earth by $20^\circ$ and $-60^\circ$ inclined with respect to ecliptic plane), TAIJIp (leading the Earth by $20^\circ$ and $+60^\circ$ inclined), TAIJIc (colocated and coplanar with LISA) with respect to the LISA mission (trailing the Earth by $20^\circ$ and $+60^\circ$ inclined). And the LISA-TAIJIm network has been identified as the most capable configuration for massive black hole binary observation. In this work, we examine the performance of three networks to the stochastic gravitational wave background (SGWB) especially for the comparison of two eligible configurations, LISA-TAIJIm and LISA-TAIJIp. This investigation shows that the detectability of LISA-TAIJIm is competitive with the LISA-TAIJIp network for some specific SGWB spectral shapes. And the capability of LISA-TAIJIm is also identical to LISA-TAIJIp to separate the SGWB components by determining the parameters of signals. Considering the performances on SGWB and massive black hole binaries observations, the TAIJIm could be recognized as an optimal option to fulfill joint observations with LISA.

\end{abstract}

\maketitle

\section{Introduction}

More than fifty gravitational wave (GW) events have been detected during the Advanced LIGO and Advanced Virgo observing runs O1-O3a, and all signals are from compact binary coalescences \cite{Abbott:2016blz,GW170817:detection,LIGOScientific:2016dsl,LIGOScientific:2018mvr,LIGOScientific:2020ibl,LIGOScientific:2021usb}. As another important targeting source, the stochastic GW background (SGWB) may encode the information about the early Universe, and the searches for SGWB from Advanced LIGO and Advanced Virgo runs are actively ongoing \cite{TheLIGOScientific:2016dpb,Abbott:2018utx,LIGOScientific:2019vic,Abbott:2021xxi,Abbott:2021ksc,Abbott:2021jel}. The NANOGrav (the North American Nanohertz Observatory for Gravitational Waves) has reported strong evidence of a stochastic process, however, no statistically significant evidence has been found yet to claim a SGWB detection \cite{NANOGrav:2020bcs}.

The space mission, LISA, is scheduled to be launched in the 2030s and targeting to detect GW in the milli-Hz frequency band \cite{2017arXiv170200786A}. The LISA interferometer is formed by three spacecraft with $2.5 \times 10^6$ km separation and trails the Earth by $20^\circ$ on a heliocentric orbit to balance the telemetry capabilities, gravity perturbation reduction, and launch vehicle \cite{2017arXiv170200786A,LISA2000}. The formation plane of the constellation is designed to be $+60^\circ$ inclined with respect to the ecliptic plane to maintain the stability of constellation. As another space GW detector, TAIJI is proposed to be a LISA-like mission with a $3 \times 10^6$ km arm length and observe the GWs in the 2030s as well \cite{Hu:2017mde}. The joint LISA-TAIJI network has been studied to bring merits for massive black hole (MBH) binary observations \cite{Ruan:2020smc,Wang:2020a,Wang:2021polar,Shuman:2021ruh}, SGWB detections \cite{Omiya:2020fvw,Seto:2020mfd,Orlando:2020oko,Pol:2021uol}, and cosmological parameter estimations \cite{Wang:2021srv,Wang:2020dkc}.

In previous work, we proposed three TAIJI orbits to construct LISA-TAIJI networks and investigated their performances on sky localizations for MBH binaries, constraints on polarizations, and overlap reduction functions \cite{Wang:alternative}. And the  LISA and different TAIJI orbital configurations are specified as follows.
\begin{itemize}
\item[(a)] LISA, which trails the Earth by $\sim 20^\circ$, and its orientation of formation plane is $+ 60^\circ$ with respect to the ecliptic plane. 
\item[(b)] TAIJIm, which leads the Earth by $\sim 20^\circ$, with a $-60^\circ$ inclined orientation.
\item[(c)] TAIJIp, which leads the Earth by $\sim 20^\circ$, with a $+60^\circ$ inclined orientation.
\item[(d)] TAIJIc, which trails the Earth by $\sim  20^\circ$ and is coplanar with LISA. 
\end{itemize}
And their orbital deployments are shown in Fig. \ref{fig:LISA_TAIJI}. Restating the previous investigation results, in three LISA-TAIJI networks (LISA-TAIJIm, LISA-TAIJIp, and LISA-TAIJIc), the LISA-TAIJIm demonstrates the best performance for MBH binary observations and shows the lower overlap reduction function between the two detectors. The detectability of LISA-TAIJIp is slightly worse than LISA-TAIJIm network for MBH binary observation, but they can have a higher overlap reduction function than LISA-TAIJIm network. The TAIJIc achieves an optimal cross-correlation with LISA, but the performance of LISA-TAIJIc is much worse than the other two networks for MBH binaries observation. On the other side, the colocated LISA and TAIJIc may be subject to similar space environments and cause correlated noises. Therefore, the TAIJIc should not be an optimal deployment for comprehensive considerations, and TAIJIm and TAIJIp would be the qualified candidates to construct a joint network with the LISA detector.

In this work, we further investigate the detectability of three LISA-TAIJI networks on the SGWB observation especially for the comparison of LISA-TAIJIm and LISA-TAIJIp. The SGWB is supposed to be composed of two kinds of sources: astrophysical origin and cosmological origin. The former one could be yielded by the overlapped GWs from abundant unresolved compact binary systems \cite{LIGOScientific:2019vic,Korol:2017qcx,Cornish:2017vip,Romano:2016dpx}. The cosmological sources could be produced by multiple mechanisms in the early Universe such as time-varying scalar fields, preheating, phase transitions, and cosmic strings, etc \cite{Caprini:2015zlo,Bartolo:2016ami,Caprini:2019egz,Caprini:2019pxz,Flauger:2020qyi,Christensen:2018iqi}. One of the targeting sources for space-borne detectors is the cosmological origin SGWB which may significantly improve our understandings of the early Universe and particle physics beyond the standard model \cite{2017arXiv170200786A}. 

To examine the capabilities of the LISA-TAIJI networks, four signal models are assumed to represent the possible SGWB spectrum. A power-law shape is employed to describe the astrophysical origin SGWB which is yielded by the unresolved BH and neutron star binaries \cite{LIGOScientific:2019vic} (stochastic foreground generated by the galactic binaries is not included). For the cosmological SGWB, three spectral shapes (flat, broken power-law, and single peaked) are selected to depict the possible SGWBs from processes in the early Universe \cite{Caprini:2015zlo,Caprini:2019egz,Caprini:2019pxz,Flauger:2020qyi}. As the first step, the sensitivity of each networks' cross-correlation to the SGWB is evaluated. The LISA-TAIJIp shows better sensitivity than LISA-TAIJIm in frequencies lower than 1 mHz because of the stronger correlation between two detectors. The LISA-TAIJIm becomes more sensitive than LISA-TAIJIp in the frequency band [1, 8] mHz. To quantify the performances, the SNRs from each pairs' cross-correlation are calculated by tuning the parameters in the signal models. The SNRs of LISA-TAIJIm are competitive to the LISA-TAIJIp network in the selected parameter spaces, and LISA-TAIJIm could surpass LISA-TAIJIp for the assumed fiducial cases. We further examine the abilities of three networks to determine the SGWB parameters and separate the astrophysical and cosmological components. Both LISA-TAIJIm and LISA-TAIJIp could improve the parameter resolutions by a factor of $\sim$1.8 compared to single LISA mission, and LISA-TAIJIc could promote this factor to $\sim$2.3-3.2. Therefore, considering the gains for MBH binaries observation, the TAIJIm could be a better choice than TAIJIp to assemble its observations with LISA.

This paper is organized as follows. 
In Sec. \ref{sec:network_SGWB_models}, we introduce three LISA-TAIJI network configurations and the assumed SGWB spectral shapes. 
In Sec. \ref{sec:cross_correlation}, we evaluate the sensitivities of each mission and their joint cross-correlation for SGWB observation, and compare the SNRs obtained from each LISA-TAIJI network for SGWB signals with different parameters.
In Sec. \ref{sec:seperate_components}, by employing Fisher matrix analysis, we examine the capability of each network to determine SGWB parameters from the combined astrophysical and cosmological signals.
We recapitulate our conclusions in Sec. \ref{sec:conclusions}. (We set $G=c=1$ in this work except otherwise stated).

\section{LISA-TAIJI networks and assumed SGWB spectral shapes} \label{sec:network_SGWB_models}

\subsection{The deployments of LISA and TAIJI}

\begin{figure}[htb]
\includegraphics[width=0.45\textwidth]{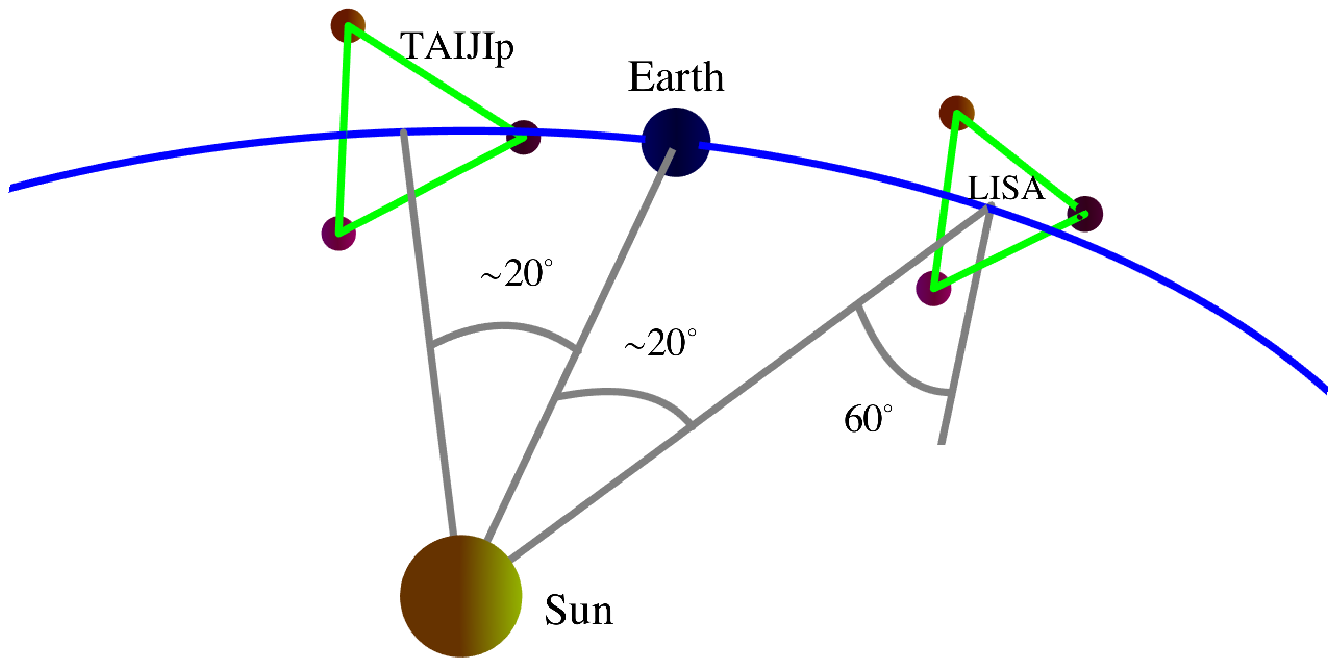}
\includegraphics[width=0.45\textwidth]{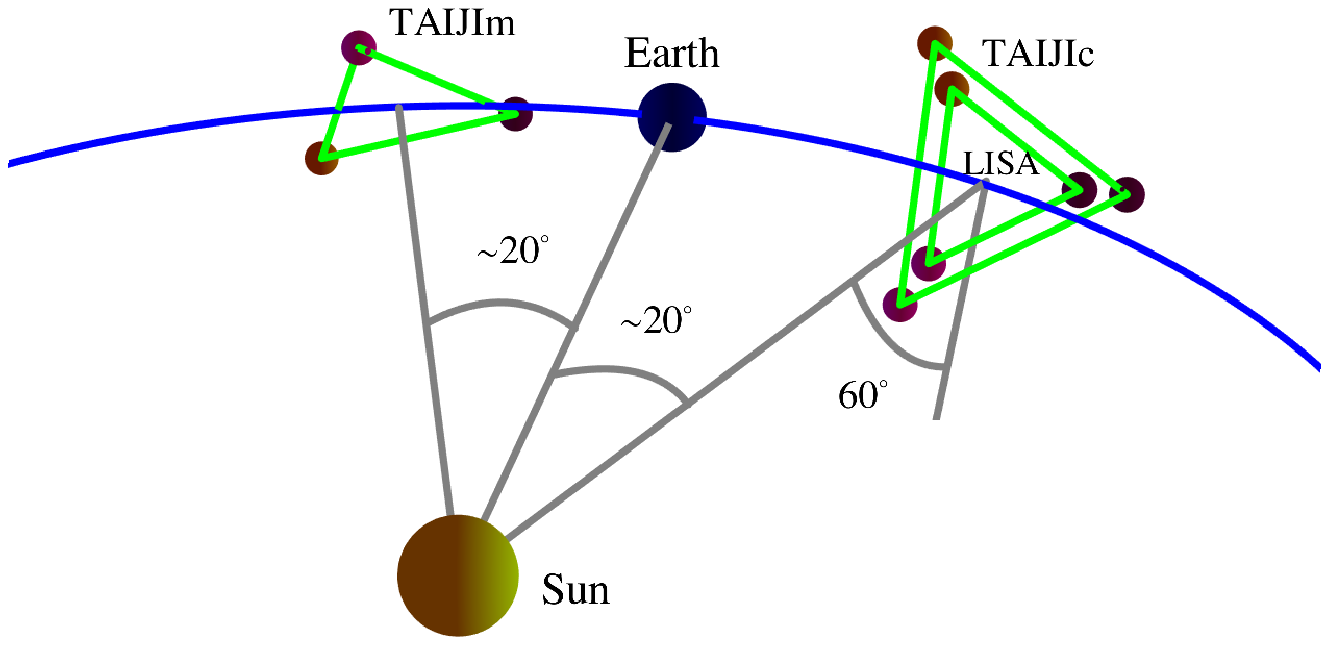}
\caption{\label{fig:LISA_TAIJI} The diagram of LISA and TAIJI missions' orbital configurations. The upper panel shows the LISA (trailing the Earth by $\sim$20$^\circ$ and $+60^\circ$ inclined with respect to the ecliptic plane) and the TAIJIp (leading the Earth by $\sim$20$^\circ$ with an inclination $+60^\circ$). The lower panel shows TAIJIm (leading the Earth by $\sim$20$^\circ$ and $-60^\circ$ inclined) and TAIJIc (colocated and coplanar with LISA). The angle between the LISA and TAIJIp formation planes is $\sim$34.5$^\circ$, and the angle for LISA and TAIJIm constellation is $\sim$71$^\circ$.
}
\end{figure}

The LISA and TAIJI are proposed space GW missions to be launched in the same epoch, and targeting to detect the GW in the frequency band 0.1 mHz - 0.1 Hz \cite{2017arXiv170200786A,Hu:2017mde}. A GW interferometer network formed by properly deployed detectors will increase the observation efficiency (for instance, SNR cumulation, sky localization, intrinsic parameter determination, and etc) comparing to single detector. In previous work \cite{Wang:alternative}, we proposed three alternative TAIJI orbits to construct networks with LISA as shown in Fig. \ref{fig:LISA_TAIJI}, and the main differences between the networks are the separation between two constellations and their orientation of the constellation plane. 

For their performances on MBH binary observations as we investigated, the large separation between LISA and TAIJIp/TAIJIm ($D_\mathrm{sep} \sim1 \times 10^8$ km) yields significant improvement on source sky localizations than on the LISA and LISA-TAIJIc network \cite{Wang:alternative}. Benefiting from the more misaligned formation plane ($\sim$71$^\circ$), LISA-TAIJIm could achieve a higher parameter resolution than LISA-TAIJIp network (with a $\sim$34.5$^\circ$ misaligned angle). On the other side, because of a larger misaligned formation plane, the overlap reduction function between LISA and TAIJIm is lower than LISA-TAIJIp especially for the frequencies lower than 1 mHz. Furthermore, the larger separated distance between two detectors lowers down a critical frequency by $f_\mathrm{crit} = c/(2D_\mathrm{sep}) = 1.5$ mHz ($c$ is speed of light) as shown in Appendix Fig. \ref{fig:overlap_reduction_fn}. To further compare the performance of the LISA-TAIJIm and LISA-TAIJIp networks, we resume the investigation to explore their detectabilities to isotropic SGWB. Although LISA-TAIJIc network is supposed to be ineligible considering its low improvement on MBH binary observation, our examinations also include it for comparison and integrity.

The LISA is a nominal 4 years mission and could be extensible to 10 years \cite{2017arXiv170200786A}.
In a realistic scenario, the observation would be interrupted by antenna repositioning and other operations, and only 75\% scientific duty cycle can be expected \cite{Caprini:2019pxz}. Therefore, only 3 years of data is suppose to be effective in the 4 years observation. In the following investigation, observation time $T_\mathrm{obs} = 3$ years is preset for LISA and TAIJI joint observation.

\subsection{Time-delay interferometry and noise budgets} \label{secsub:TDI}

Time-delay interferometry (TDI) will be employed for both LISA and TAIJI to suppress the laser frequency noise and achieve targeting sensitivity. The principle of the TDI is to combine multiple time-shifted interferometric links and obtain an equivalent equal path for two interferometric laser beams. The GW response of TDI is combined by the response of every single link, and formulation has been specified in \cite{1975GReGr...6..439E,1987GReGr..19.1101W,Vallisneri:2007xa,Vallisneri:2012np,Tinto:2010hz}. 

For a LISA-like with six laser links, three optimal TDI channels (A, E, T) could be constructed from three first-generation Michelson TDI configuration (X, Y, Z),
\begin{equation} \label{eq:optimalTDI}
 {\rm A} =  \frac{ {\rm Z} - {\rm X} }{\sqrt{2}} , \quad {\rm E} = \frac{ {\rm X} - 2 {\rm Y} + {\rm Z} }{\sqrt{6}} , \quad {\rm T} = \frac{ {\rm X} + {\rm Y} + {\rm Z} }{\sqrt{3}},
\end{equation}
where Y and Z are obtained by circulating of the spacecraft indexes in the X channel, and the Michelson-X channel is
\begin{equation} \label{eq:X_measurement}
\begin{aligned}
{\rm X} =& [ \mathcal{D}_{31} \mathcal{D}_{13} \mathcal{D}_{21} \eta_{12}  + \mathcal{D}_{31}  \mathcal{D}_{13} \eta_{21}  +  \mathcal{D}_{31} \eta_{13} +  \eta_{31}   ] \\
& - [ \eta_{21} + \mathcal{D}_{21} \eta_{12} +\mathcal{D}_{21} \mathcal{D}_{12} \eta_{31} + \mathcal{D}_{21}  \mathcal{D}_{12} \mathcal{D}_{31} \eta_{13} ], \\
\end{aligned}
\end{equation}
where $\mathcal{D}_{ij}$ is a time-delay operator, $ \mathcal{D}_{ij} \eta(t) = \eta(t - L_{ij} )$, $\eta_{ji}$ are the Doppler measurement from S/C$j$ to S/C$i$. By adopting new designs in \cite{Otto:2012dk,Otto:2015,Tinto:2018kij}, and the expression of measurement $\eta_{ij}$ composited by GW signal and noises are referred to our recent work \cite{Wang:2ndTDI,Wang:2021polar}. And the specific formulation for GW response are reiterated in Appendix \ref{sec:appendix_response}. 
These joint optimal channels represent the eventual detectability of a mission \cite{Prince:2002hp,Vallisneri:2007xa}.

By assuming laser frequency noise is sufficiently suppressed in TDI, the acceleration noise and optical path noise are considered to be the dominant noises for GW observation, and they are included to evaluate the sensitivity of a mission to SGWB. The budges of acceleration noise for LISA and TAIJI are treated as same \cite{2017arXiv170200786A,Luo:2020},
\begin{equation}
 S_{\rm acc} = 9 \ \frac{\rm fm^2/s^4}{ \rm Hz } \left[ 1 + \left( \frac{0.4 \ {\rm mHz}}{f} \right)^2 \right]  
 \left[ 1 + \left(\frac{f}{8 \ {\rm mHz}} \right)^4 \right] .
\end{equation}
And their optical path noise budgets are slightly different as
\begin{align}
 S_{\rm op, LISA} & = 100 \ \frac{\rm pm^2}{\rm Hz} \left[ 1 + \left(\frac{2 \ {\rm mHz}}{f} \right)^4 \right],  \\
S_{\rm op, TAIJI} & = 64 \ \frac{\rm pm^2}{\rm Hz} \left[ 1 + \left(\frac{2 \ {\rm mHz}}{f} \right)^4 \right].
 \end{align}
The power spectrum density (PSD) of noise in a TDI channel $\mathrm{S}_\mathrm{n,TDI}$ is calculated by using the numerical method, and we refer to our previous works for detailed algorithm \cite{Wang:1stTDI,Wang:2ndTDI}.

\subsection{SGWB models} \label{subsec:SGWB_model}

The isotropic SGWB could be characterized as the variation of the energy density in the frequency domain,
\begin{equation}
 \Omega_\mathrm{GW} = \frac{1}{\rho_c} \frac{\mathrm{d} \rho_\mathrm{GW} }{ \mathrm{d} \ln f },
\end{equation}
where $\rho_c = \frac{3 H^2_0 c^2}{8 \pi G}$ is the critical density of the Universe \cite{Allen:1996vm,Allen:1997ad}, and $H_0 \simeq 2.185 \times 10^{-18}$ Hz is the Hubble constant \cite{Planck:2018vyg}. The SGWB could be yields by astrophysical sources and cosmological mechanisms, and kinds of cosmological SGWB signals are predicated during the processes of the early Universe \cite{Caprini:2015zlo,Caprini:2019egz,Kuroyanagi:2018csn}. Four SGWB spectral shapes are assumed to perform our investigation as follows \cite{Caprini:2019pxz,Flauger:2020qyi,Martinovic:2020hru}.
\begin{itemize}
\item[1)] Power-law (PL) SGWB signal,
\begin{equation}
\Omega_\mathrm{PL} = \Omega_{0} \left( \frac{f}{f_\mathrm{ref}} \right)^{\alpha_0}, \label{eq:PL_signal} 
\end{equation}
where $\Omega_{0}$ is the amplitude of the SGWB energy density, and $\alpha_{0}$ is the index of the power law. This power law background is expected to be generated by abundant unresolved astrophysical BH and NS binaries. The fiducial signal is selected to be $\Omega_0 = 4.446 \times 10^{-12}$ and $\alpha_0 = 2/3$ at reference frequency $f_\mathrm{ref} = 1 \ \mathrm{mHz} $ \cite{LIGOScientific:2019vic}.

\item[2)] Flat SGWB signal,
\begin{equation}
\Omega_\mathrm{flat} = \Omega_1. \label{eq:flat_signal}
\end{equation}
The flat SGWB is a simply assumed cosmological SGWB which permeates all frequency range with a constant amplitude \cite{Cornish:2001qi}. And we choose $\Omega_1 = 1 \times 10^{-11} $ as the fiducial value for this shape.

\item[3)]  Broken power-law (BPL) SGWB signal,
\begin{equation}
 \Omega_{\rm BPL} = \Omega_1 \left( \frac{f}{ f_\mathrm{ref} } \right)^{\alpha_1} \left[ 1 + 0.75 \left( \frac{f}{ f_\mathrm{ref} } \right)^\Delta \right]^{ (\alpha_2 - \alpha_1 ) / \Delta }. \label{eq:BPL_signal}
\end{equation}
The broken power-law SGWB may produced during the first-order phase transition \cite[and reference therein]{Hindmarsh:2013xza,Caprini:2015zlo,Caprini:2019egz,Martinovic:2020hru}. The fiducial parameters are presumed to be $ \alpha_1 = 3, \alpha_2 = -4$, and $\Delta =2 $, and the amplitude $\Omega_1 = 1 \times 10^{-9}$ and reference frequency $ f_\mathrm{ref} = 10 \ \mathrm{mHz}$.

\item[4)]  Single peaked (SP) SGWB signal,
\begin{equation}
 \Omega_{\rm SP} = \Omega_1 \exp\left[ - \frac{ \left( \log_{10} ( f /f_\mathrm{ref} ) \right)^2}{ \Delta^2 } \right]. \label{eq:SP_signal}
\end{equation}
A single peaked SGWB signal could be a cosmological source and yielded by during particular processes in the early Universe, and the typical values of parameters are set to be $\Delta = 0.2 $, $\Omega_1 = 1 \times 10^{-11}$, and $f_\mathrm{ref} = 3 \ \mathrm{mHz}$ \cite{Caprini:2019pxz,Flauger:2020qyi}.
\end{itemize}
The energy spectral densities of four SGWB models with their respective fiducial values are plotted in the upper panel of Fig. \ref{fig:Omega_sensitivity}. And the conversion from relative energy density to PSD will be
\begin{equation} 
P_h (f)  = \frac{3 H^2_0}{4 \pi^2 f^3} \Omega_\mathrm{GW} (f). \label{eq:P_h}
\end{equation}
For the LISA-like mission, considering the antenna pattern of an interferometer, the PSD of SGWB in a TDI channel is expected to be 
\begin{equation}
S_{h, \rm TDI} (f) = P_h (f)  \mathcal{R}_\mathrm{TDI} (f), \label{eq:S_h}
\end{equation}
where $ \mathcal{R}_\mathrm{TDI} $ is the averaged response function of TDI channel in the frequency domain,
\begin{equation} 
\begin{aligned}
 \mathcal{R}_{\rm TDI} (f) =& \frac{1}{4 \pi^2}  \int^{2 \pi}_{0} \int^{\frac{\pi}{2}}_{-\frac{\pi}{2}} \int^{\pi}_{0} |F^h_{ \rm TDI} (f, \mathbf{n})|^2 \cos \beta {\rm d} \psi {\rm d} \beta {\rm d} \lambda,
\end{aligned}
\end{equation}
and $F^h_{ \rm TDI}$ is calculated by using Eqs. \eqref{eq:optimalTDI} and \eqref{eq:TDI_Fh}.

\section{Detecting SGWB with cross-correlation between LISA and TAIJI} \label{sec:cross_correlation}

In this section, we will investigate and compare the detectability of LISA-TAIJI configurations for four kinds of SGWB models. Similar to the ground-based GW detectors, the sensitivity of the LISA-TAIJI networks to the stochastic signals will depend on the cross-correlation between the interferometers. 

\subsection{The sensitivities of LISA-TAIJI networks}

The cross-correlation of data streams from two interferometers in Fourier domain could be described as \cite{Allen:1996vm,Allen:1997ad,Cornish:2001qi},
\begin{equation}
C \simeq \int^{+\infty}_{-\infty} \mathrm{d} f  \int^{+\infty}_{-\infty} \mathrm{d} f^\prime \delta_T (f - f^\prime) \tilde{d}_i (f)  \tilde{d}^\ast_j (f^\prime) \tilde{Q}( f^\prime ), 
\end{equation}
where $\tilde{ d }_i = \tilde{s}_i + \tilde{n}_i $ is the data in TDI $i$ channel, $\tilde{s}_i$ is the SGWB signal, $\tilde{n}_i$ the detector noise, $\delta_T = \frac{\sin(\pi f T_\mathrm{obs})}{\pi f}$, and $ \tilde{Q}$ is a filter function. Assuming the noises are not correlated between two TDI channels from different missions, and there is also no correlation between noises and signal, the average of $C$ will be
\begin{equation} \label{eq:C_average}
\begin{aligned}
 \left\langle C \right\rangle & = \left\langle s_i, s_j \right\rangle + \left\langle s_i , n_j \right\rangle  + \left\langle n_i , s_j \right\rangle + \left\langle n_i , n_j \right\rangle \\
 & = \left\langle s_i , s_j \right\rangle \\
 & = \int^{+\infty}_{-\infty} \mathrm{d} f  \int^{+\infty}_{-\infty} \mathrm{d} f^\prime \delta_T (f - f^\prime) \left\langle \tilde{d}_i (f)   \tilde{d}^\ast_j (f^\prime) \right\rangle \tilde{Q}( f^\prime ) \\
 & = \frac{T_\mathrm{obs} }{2}  \int^{+\infty}_{-\infty} \mathrm{d} f  P_h (f) \gamma_{ij} (f) \tilde{Q}( f ).
 \end{aligned}
\end{equation}
The correlated SGWB signal in two TDI channels is implemented in the last row of Eq. \eqref{eq:C_average}, 
\begin{equation}
 \left\langle \tilde{d}_i (f)  \tilde{d}^\ast_j (f^\prime) \right\rangle = \frac{1}{2} \delta_T ( f - f^\prime ) P_h (f) \gamma_{ij} (f),
\end{equation}
where $\gamma_{ij}$ is the overlap reduction function between LISA TDI $i$ channel and TAIJI TDI $j$ channel, 
\begin{equation}
\gamma_{ij} (f) = \frac{1}{4 \pi} \int  \mathrm{d}^2 \mathbf{n}  F_{i, \rm LISA} (f, \mathbf{n})  F_{j, \rm TAIJI} (f, \mathbf{n}),  \label{eq:gamma_ij} 
\end{equation}
where $F_{i, \mathrm{LISA/TAIJI} }$ is the response function from TDI channel $i$ ($i$=A/E/T) of LISA/TAIJI as calculated by Eqs. \eqref{eq:optimalTDI} and \eqref{eq:TDI_Fh}. The overlap reduction functions for each LISA-TAIJI pair have been calculated in previous work \cite{Wang:alternative}, and the normalized overlap reduction functions of three LISA-TAIJI networks are plotted in Fig. \ref{fig:overlap_reduction_fn}.

The SNR $\rho$ from two detector cross-correlations are expected to be \cite[and references therein]{Allen:1996vm,Allen:1997ad,Cornish:2001qi,Romano:2016dpx}, 
\begin{equation}
\rho^2  \equiv \frac{  \left\langle C \right\rangle^2 }{ \left\langle N^2 \right\rangle} = \frac{  \left\langle C \right\rangle^2 }{ \left\langle C^2 \right\rangle - \left\langle C \right\rangle^2 }.
\end{equation}
And the variance is
\begin{equation}
\begin{aligned}
\left\langle N^2 \right\rangle = & \left\langle C^2 \right\rangle - \left\langle C \right\rangle^2 \\
= &  \frac{T_\mathrm{obs} }{4}  \int^{+\infty}_{-\infty} \mathrm{d} f  M (f)  | \tilde{Q}( f ) |^2,
\end{aligned}
\end{equation}
by implementing the power spectral density of noise in TDI channel,
\begin{equation}
\left\langle \tilde{n}_i (f)   \tilde{n}^\ast_i (f^\prime) \right\rangle = \frac{1}{2} \delta_T ( f - f^\prime ) S_n (f),
\end{equation}
where
\begin{equation}
\begin{aligned}
M = & S_{n, i} (f) S_{n,j} (f) + S_{n, i} (f) S_{h,j} (f) + S_{h, i} (f) S_{n,j} (f) \\
 & + S_{h, i} (f) S_{h,j} (f) + \gamma^2_{ij} (f) P^2_h (f).
\end{aligned}
\end{equation}
By defining the inner product
\begin{equation}
 ( A , B ) = \int^{+\infty}_{-\infty} \mathrm{d} f A (f) B^\ast (f) M (f) ,
\end{equation}
the $\rho^2$ will be
\begin{equation}
 \rho^2 = T_\mathrm{obs} \frac{ ( \tilde{Q} , \frac{  P_h (f) \gamma_{ij} (f) }{ M(f) }  )^2 }{ ( \tilde{Q} , \tilde{Q} ) },
\end{equation}
and then an optimal filter $\tilde{Q}( f )$ is applied to maximize the SNR,
\begin{equation}
 \tilde{Q}( f ) = \frac{  P_h (f) \gamma_{ij} (f) }{ M(f) }.
\end{equation}
Therefore, the SNR for SGWB observation from LISA-TAIJI correlation will be 
\begin{equation}  \label{eq:rho2_origin}
 \rho^2 =  T_\mathrm{obs} \int^{+\infty}_{-\infty} \mathrm{d} f \frac{ \gamma^2_{ij} (f)  P^2_h (f)  }{ M(f) }
\end{equation}
For a weak-signal ($S_{n, i} \gg S_{h,i} $), the SNR could be approximated as
\begin{equation} \label{eq:rho2}
\begin{aligned}
 \rho^2 \simeq & \sum_{i,j = \mathrm{A, E, T} }  T_\mathrm{obs} \int^{+\infty}_{-\infty} \mathrm{d} f \frac{ \gamma^2_{ij} (f)  P^2_h (f)  }{S_{n, i} (f) S_{n,j} (f) } \\
 = &  \sum_{i,j = \mathrm{A, E, T} } 2 T_\mathrm{obs} \int^{+ \infty }_{0} \mathrm{d} f  \frac{ \gamma^2_{ij} (f)  P^2_h (f)}{ S_{n, i} (f)  S_{n, j} (f) } , 
\end{aligned}
\end{equation}
where $T_\mathrm{obs} = 3$ years is the observation time. 

\begin{figure}[htb]
\includegraphics[width=0.45\textwidth]{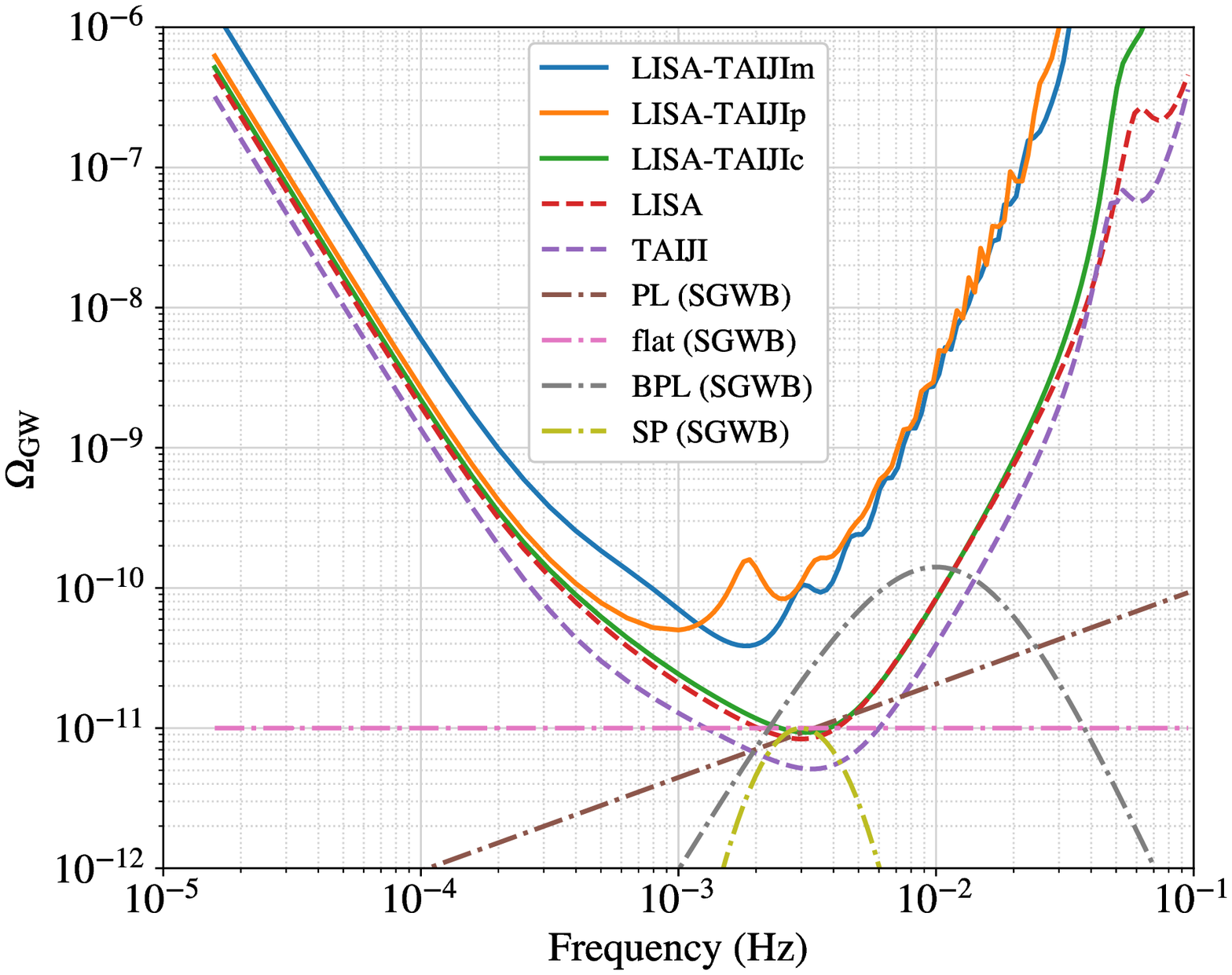} 
\includegraphics[width=0.45\textwidth]{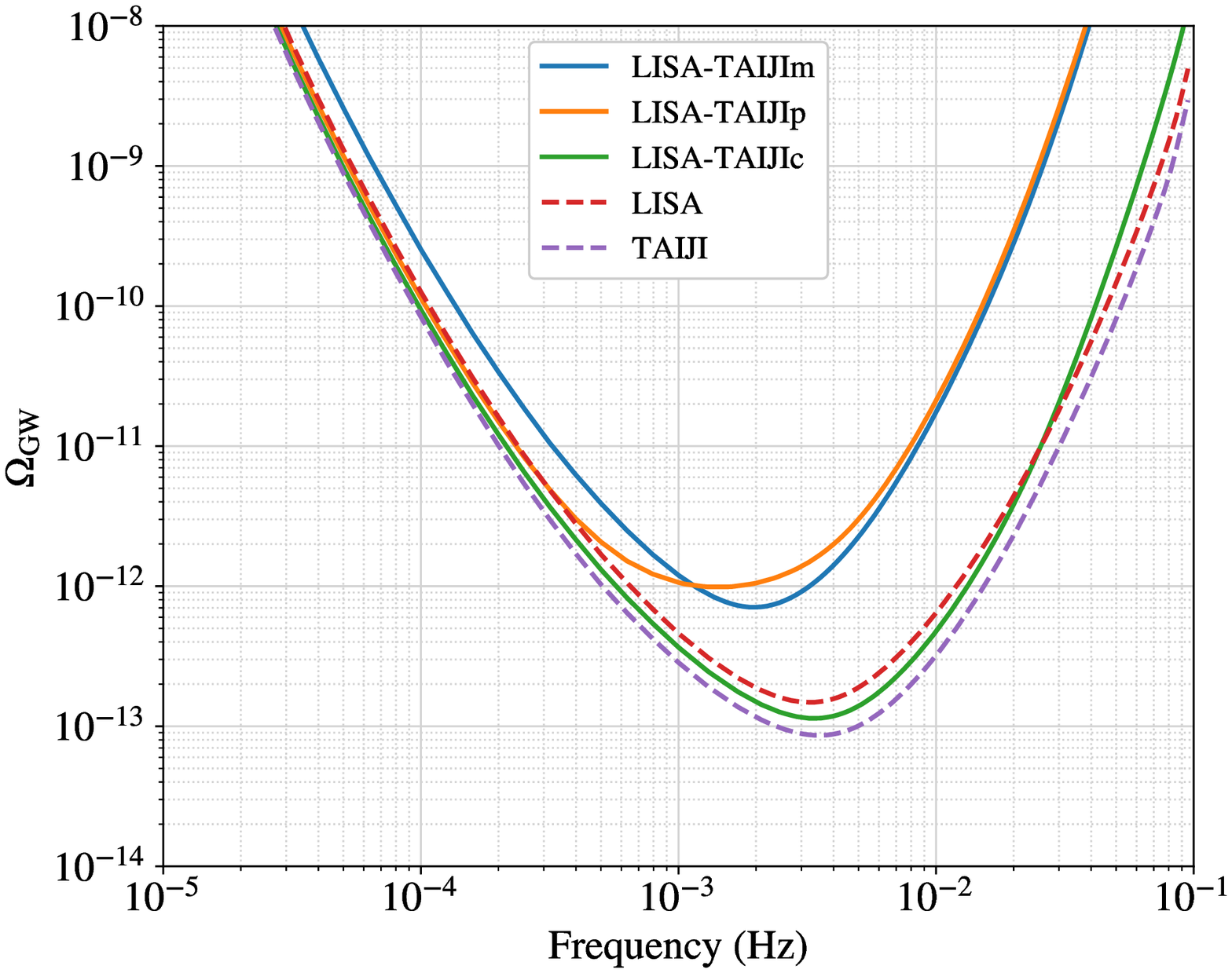} 
\caption{The assumed SGWB signals and sensitivities (upper plane) and power-law integrated sensitivities in 3 years for $\rho_\mathrm{th}= 10$ (lower panel) for LISA/TAIJI/LISA-TAIJI. The sensitivity from LISA-TAIJIp cross-correlation is better than LISA-TAIJIm for the frequencies lower than 1 mHz, and LISA-TAIJIm is more sensitive than LISA-TAIJIp for frequency band [1, 8] mHz. LISA-TAIJIc could achieve the best sensitivity in three networks. \label{fig:Omega_sensitivity} 
}
\end{figure}

To characterize the sensitivity of cross-correlation to SGWB, the equivalent energy density could be evaluated as
\begin{align}
\Omega_{\rm cross} (f) & = \frac{4 \pi^2 f^3}{3 H^2_0}   \left( \sum_{i, j = \rm A,E,T} \frac{|\gamma_{ij} (f) |^2}{ \mathrm{S}^{\rm LISA}_{ n, i} (f) \ \mathrm{S}^{\rm TAIJI}_{ n, j} (f) }   \right)^{-1/2}. \label{eq:Omega_LISA_TAIJI_sensitivty} 
\end{align}
The sensitivities of three LISA-TAIJI networks are shown by the solid curves in the upper panel of Fig. \ref{fig:Omega_sensitivity}. As we can expect, the LISA-TAIJIc shows an optimal sensitivity in the three networks. For the LISA-TAIJIm and LISA-TAIJIp pairs, subjecting to the critical frequency and instrument noises, their most sensitive frequency is around $f_\mathrm{crit} = 1.5 $ mHz. The LISA-TAIJIp is more sensitive than LISA-TAIJIm for frequencies lower than 1 mHz, while LISA-TAIJIm becomes more sensitive to SGWB in the frequency band $\sim$[1, 8] mHz.  

For a single LISA-like mission with three optimal TDI channels, the equivalent energy density of noise could be evaluated as,
\begin{align}
\Omega_{\rm mission} (f) =   \frac{4 \pi^2 f^3}{3 H^2_0}  \left( \sum_{i = \rm A,E,T} \frac{ \mathcal{R}_i (f) }{ \mathrm{S}^{\rm mission}_{n, i} (f) }  \right)^{-1} \label{eq:Omega_mission_sensitivity},
\end{align}
where $\mathcal{R}_i$ is the averaged response function of TDI $i$ channel. The corresponding sensitivities of LISA and TAIJI mission are shown by dashed curves in the upper panel of Fig. \ref{fig:Omega_sensitivity}.

To illustrate the detectability of a detector to a power-law SGWB signal, $\Omega_h = \Omega_i (f/f_\mathrm{ref})^{\alpha_i}$, a power-law integrated sensitivity is proposed in \cite{Thrane:2013oya}. Based on given observation time $T_\mathrm{obs}$ and SNR threshold $\rho_\mathrm{th}$, the power-law sensitivity (PLS) is calculated as
\begin{equation} \label{eq:Omega_PLS} 
\begin{aligned}
\Omega_{i}  = & \frac{\rho_\mathrm{th}}{\sqrt{2 T_\mathrm{obs}}} \frac{4 \pi^2 }{3 H^2_0}   \left( \sum_{i, j = \rm A,E,T} \int^{f_\mathrm{max}}_0 \mathrm{d} f \frac{|\gamma_{ij} |^2 (f/f_\mathrm{ref})^{2 \alpha_i} }{ f^6 \mathrm{S}^{\rm LISA}_{ n, i} \ \mathrm{S}^{\rm TAIJI}_{ n, j} }   \right)^{-1/2}, \\
 \Omega_\mathrm{PLS} & = \max _{\alpha_i} \left[ \Omega_i \left( \frac{f}{f_\mathrm{ref}} \right)^{\alpha_i} \right], \quad \mathrm{for} \ \alpha_i \in [-8,  8].
\end{aligned}
\end{equation}
By assuming $\rho_\mathrm{th} =  10$, the $\Omega_\mathrm{PLS} $ for each LISA-TAIJI networks are shown in Fig. \ref{fig:Omega_sensitivity} lower panel, as well as the PLS curves for LISA and TAIJI mission. As we can see, for a power-law SGWB signal, the LISA-TAIJIp will have better sensitivity than LISA-TAIJIm for the frequencies lower than 1 mHz, and LISA-TAIJIm will be more sensitive for the frequency band [1, 8] mHz.

\subsection{SNR to SGWB}

For assumed SGWB model listed by Eqs. \eqref{eq:PL_signal}-\eqref{eq:SP_signal}, the SNR is calculated by implementing Eq. \eqref{eq:rho2_origin} to evaluate the detectability of each LISA-TAIJI network. The motivation to use Eq. \eqref{eq:rho2_origin} instead of Eq. \eqref{eq:rho2} is that weak signal approximation becomes inaccurate when the amplitude of energy density of SGWB is large compared to detector noise. To compare the detectability of networks in different SGWB parameter spaces, one or two parameters in Eqs \eqref{eq:PL_signal}-\eqref{eq:SP_signal} are selected to be tunable.

\begin{figure}[htb]
\includegraphics[width=0.47\textwidth]{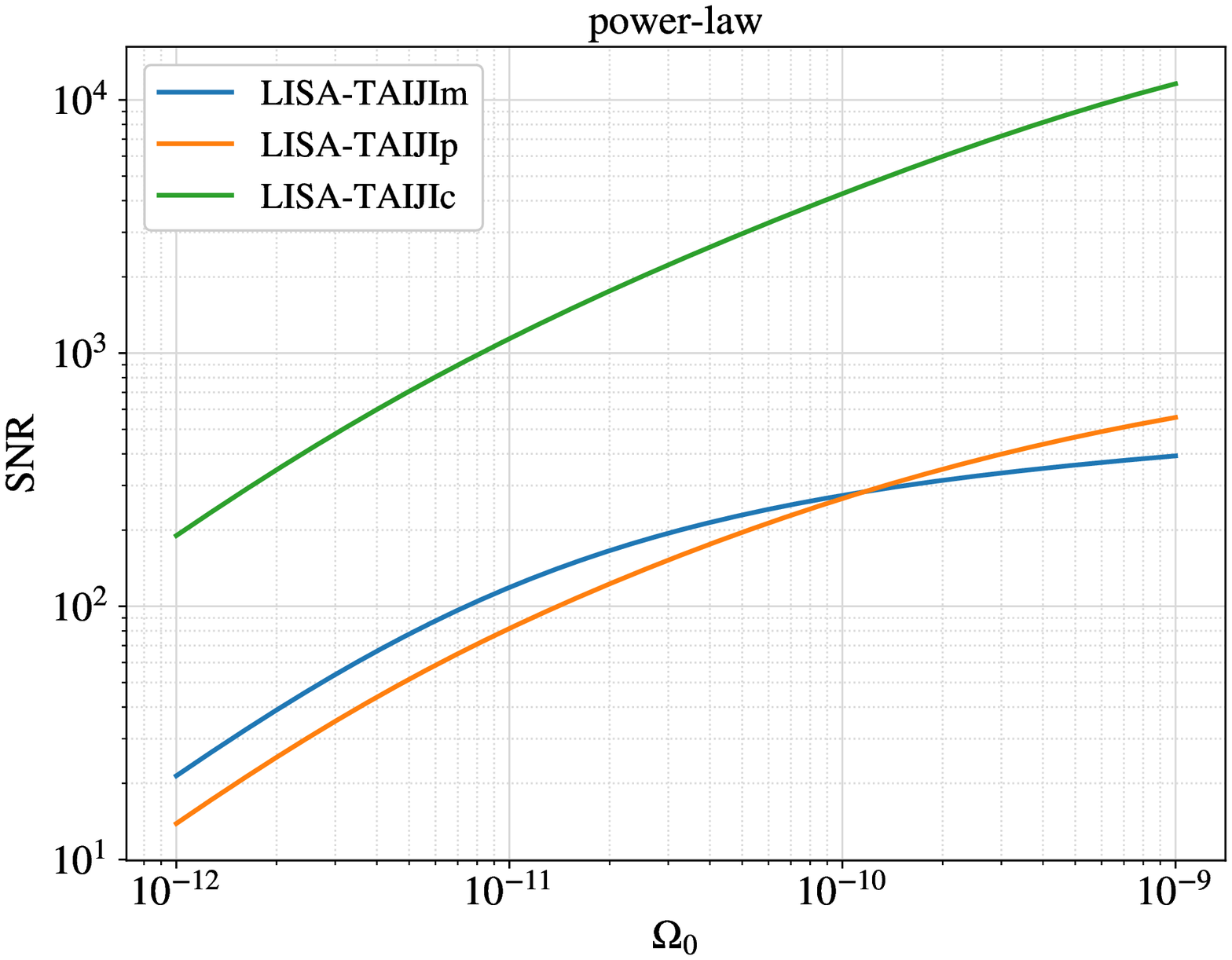} 
\includegraphics[width=0.47\textwidth]{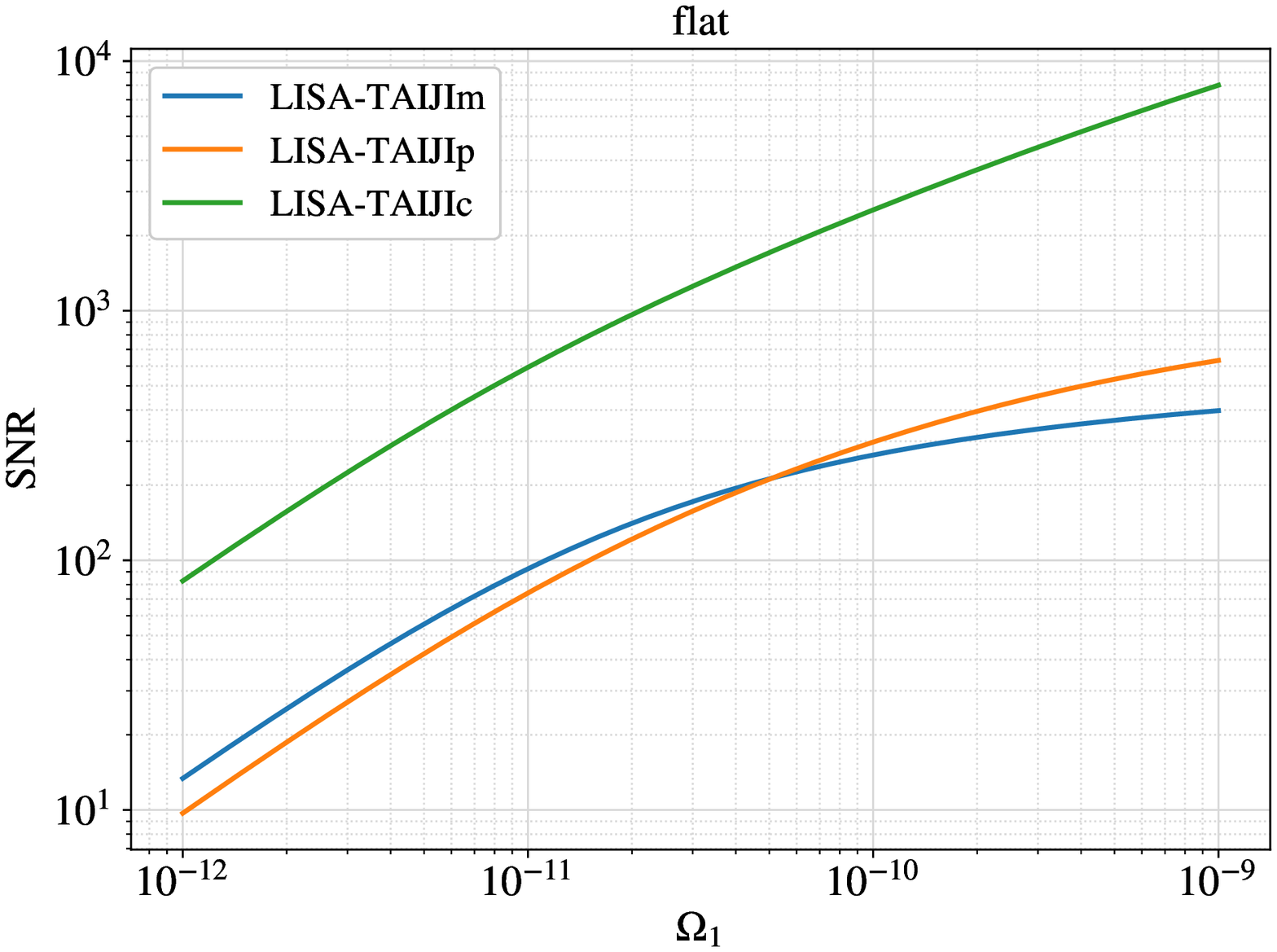} 
\caption{The SNRs from three LISA-TAIJI networks varying with amplitudes $\Omega_0$ for power-law (upper panel) and $\Omega_1$ for flat (lower panel) SGWB spectral shapes. Compared to LISA-TAIJIp configuration, LISA-TAIJIm achieves higher SNR when the amplitude of power-law SGWB $\Omega_0$ is smaller than $10^{-10}$ and amplitude of flat SGWB $\Omega_1$ is less than $5 \times 10^{-11}$. \label{fig:SNR_PL_flat} 
}
\end{figure}

For a power-law SGWB signal with a fixed power index $\alpha_0 = 2/3$, the SNR of three LISA-TAIJI networks is calculated with varying amplitude $\Omega_0$ in Eq. \eqref{eq:PL_signal}, and the results are shown in the upper plot of Fig. \ref{fig:SNR_PL_flat}. The LISA-TAIJIc network has the best capability in three networks as expected, and its SNR is almost one order higher than the other two networks. The LISA-TAIJIm shows a better detectability than LISA-TAIJIp when $\Omega_0$ is smaller than $10^{-10}$, and LISA-TAIJIp will obtain a higher SNR for $\Omega > 10^{-10}$. For the fiducial value $\Omega_0 = 4.446 \times 10^{-12}$, the LISA-TAIJIm would be a better option to detect this astrophysical SGWB.

The SNRs of three LISA-TAIJI networks varying with amplitude $\Omega_1$ of flat SGWB signal are shown in Eq. \eqref{eq:flat_signal}. The flat signal is actually a special power-law with a power index equal to zero. Similar to the power-law case, the SNR from LISA-TAIJIm is higher than LISA-TAIJIp for $\Omega < 5 \times 10^{-11}$, and SNR of LISA-TAIJIp surpass LISA-TAIJIm's for $\Omega > 5 \times 10^{-11}$.

\begin{figure*}[htb]
\includegraphics[width=0.46\textwidth]{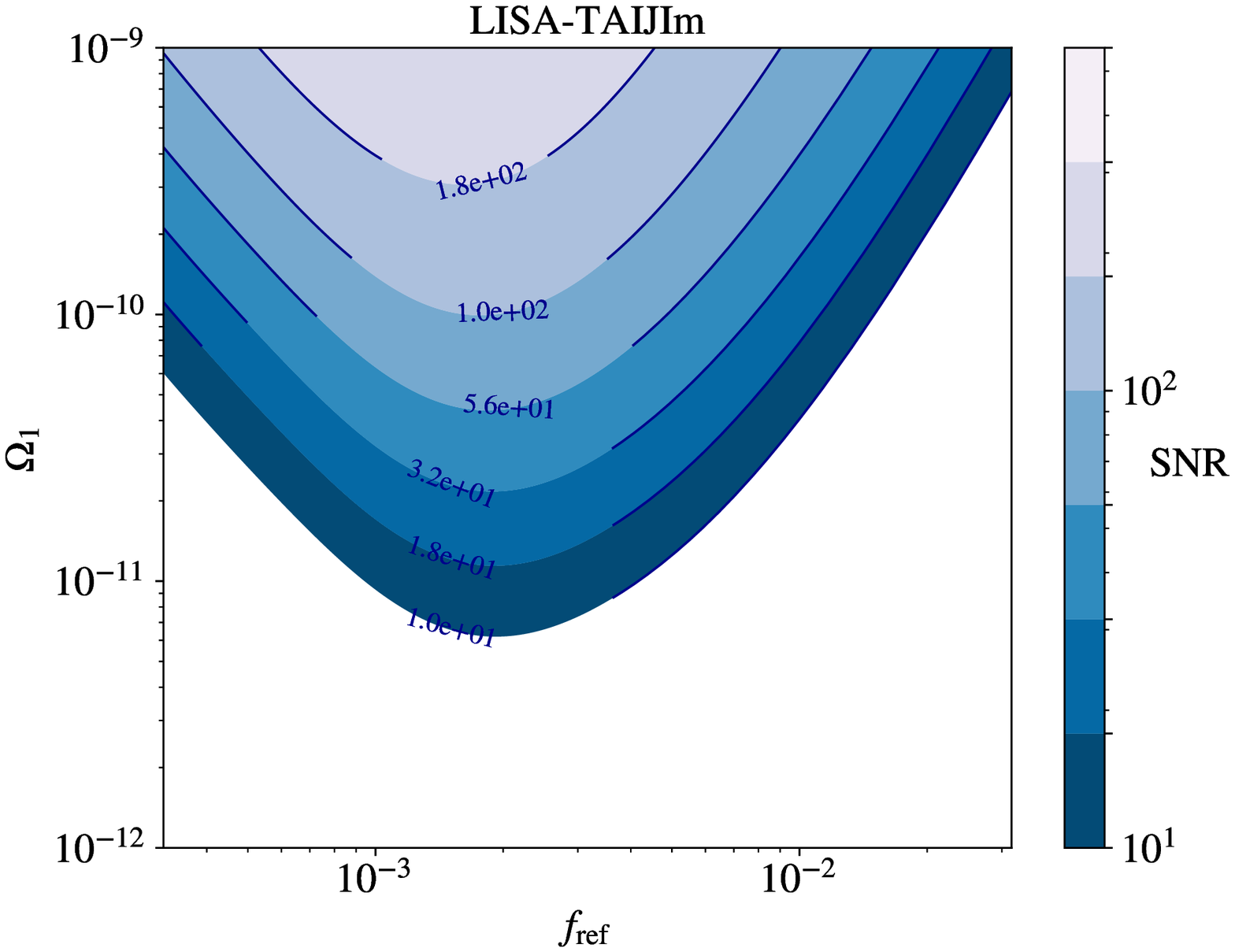} 
\includegraphics[width=0.46\textwidth]{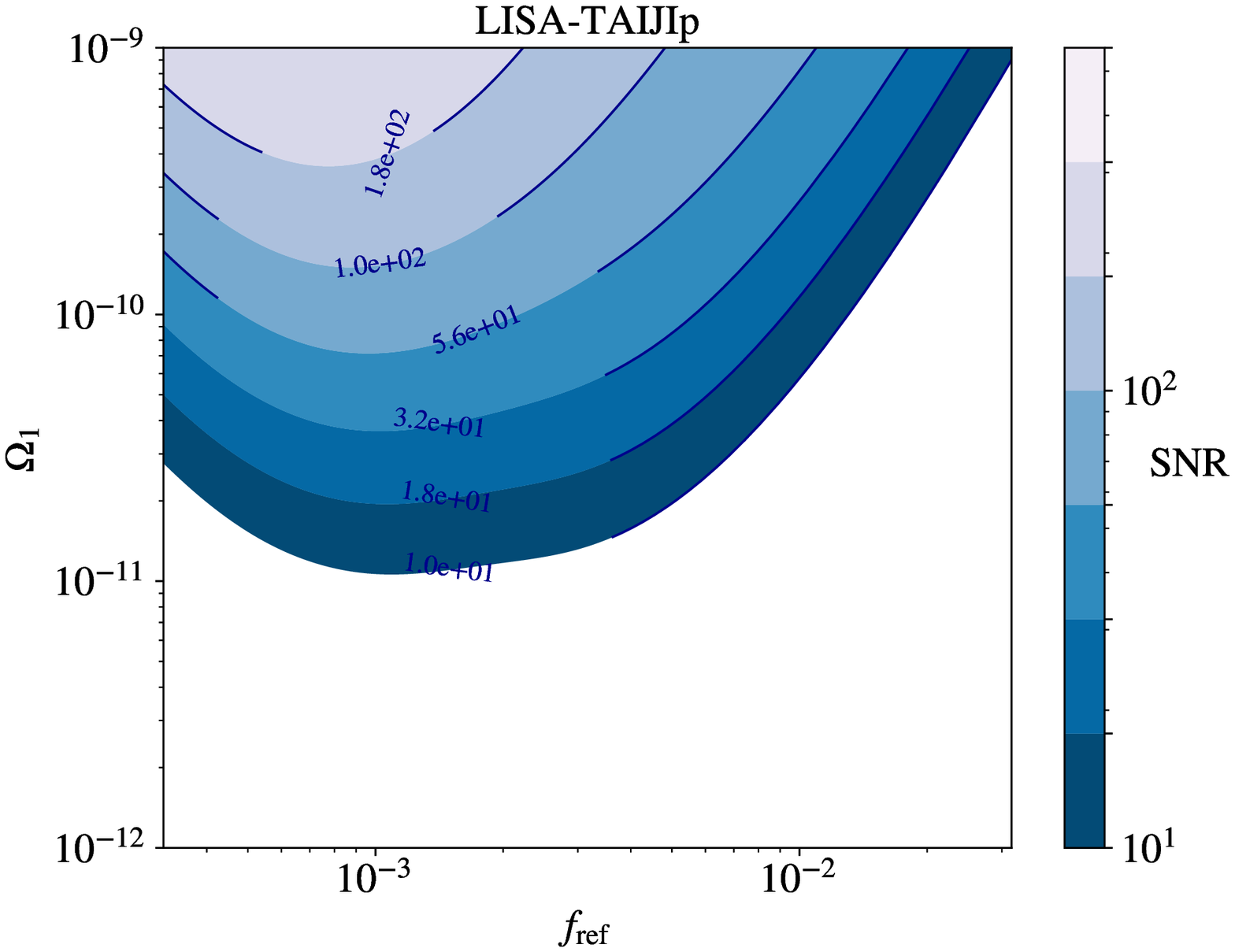} 
\includegraphics[width=0.46\textwidth]{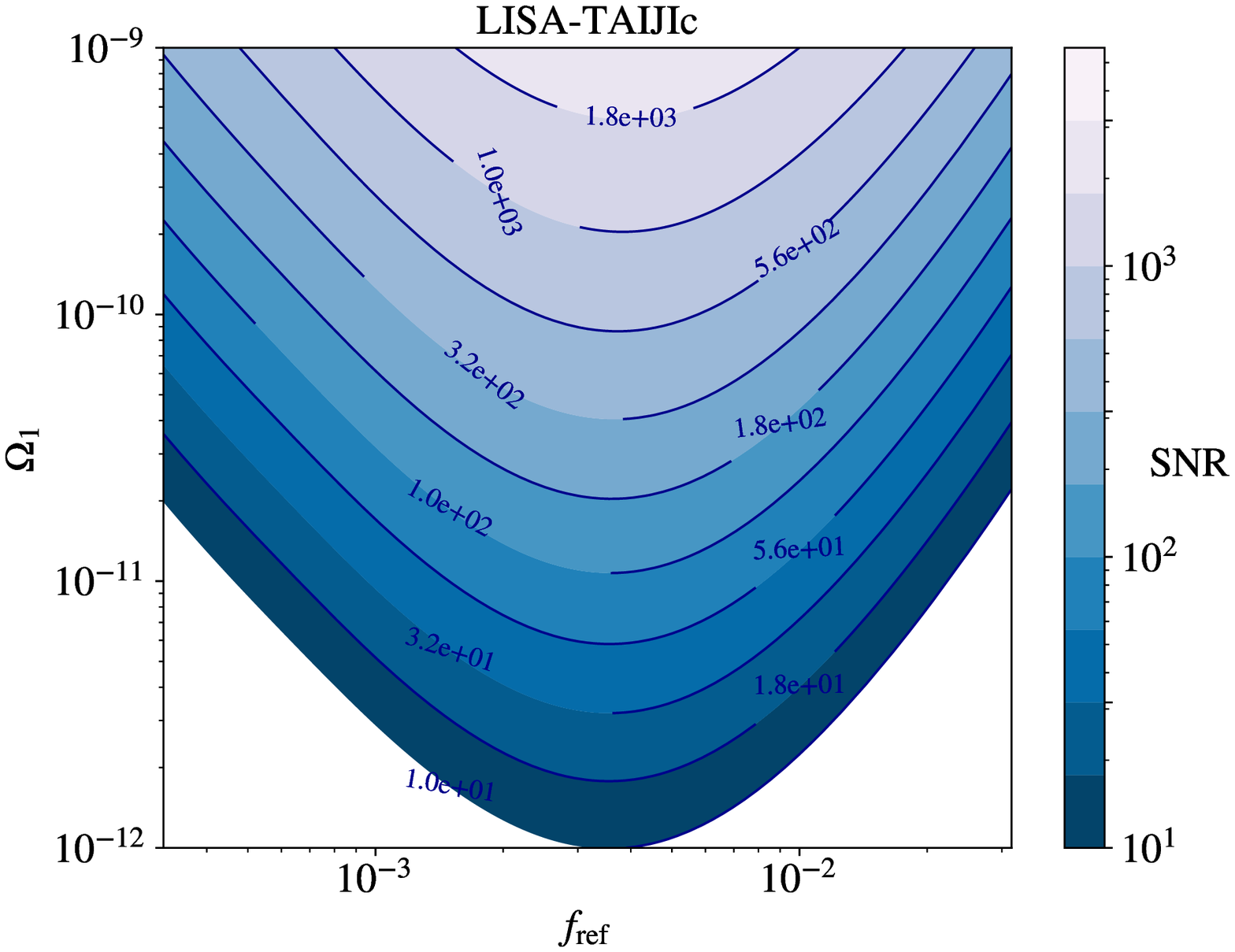} 
\includegraphics[width=0.46\textwidth]{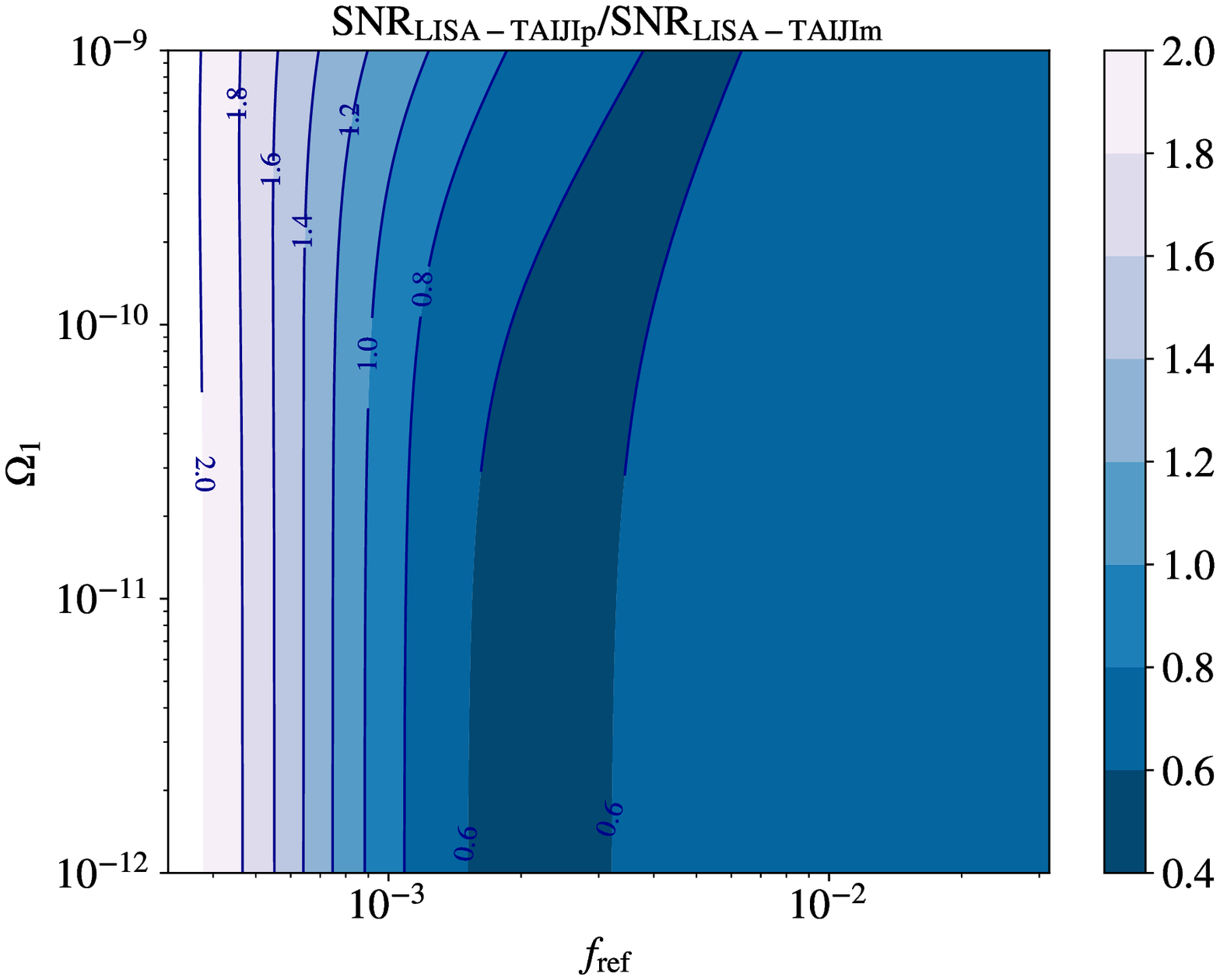} 
\caption{The SNR from LISA and TAIJI cross-correlations for broken power-law SGWB signal with different amplitude $\Omega_1$ and reference frequency $f_\mathrm{ref}$. The SNR contours for three LISA-TAIJI networks are shown in upper and lower left plots. As the SNR ratios are shown in the lower right plot, the correlation between LISA and TAIJIm would yield higher SNR than LISA-TAIJIp for $f_\mathrm{ref} \gtrsim 0.9$ mHz, and LISA-TAIJIp configuration becomes more sensitive to broken power-law signals for reference frequency lower than 0.9 mHz. \label{fig:SNR_BPL} 
}
\end{figure*}

For a broken power-law SGWB signal, the amplitude $\Omega_1$ and reference frequency $f_\mathrm{ref}$ in Eq. \eqref{eq:BPL_signal} are selected to be variables to calculate SNRs of LISA-TAIJI networks, and the parameter space are chosen to be $ \Omega_1 \in [10^{-12}$, $10^{-9}]$ and $ f_\mathrm{ref} \in [0.3, 30]$ mHz. Other parameters are fixed to be their fiducial values, ($ \alpha_1 = 3, \alpha_2 = -4$, and $\Delta =2 $). As the contour plot shown by upper left panel of Fig. \ref{fig:SNR_BPL} shows, LISA-TAIJIm could achieve its highest SNR around $f_\mathrm{ref} \simeq 2$ mHz at a given $\Omega_1$.  And LISA-TAIJIp has the highest SNR at $f_\mathrm{ref} \simeq 1$ mHz at a fixed amplitude as the upper right plot shown. The optimal reference frequency moves to 4 mHz for LISA-TAIJIc network as shown in the lower left panel, and the SNR is significantly larger than the other two networks. To compare the detectability of LISA-TAIJIm and LISA-TAIJIp in the selected parameter space, their SNR ratio is made in the lower right panel. The LISA-TAIJIm could achieve higher SNR than LISA-TAIJIp in the region $f_\mathrm{ref} \gtrsim 0.9$ mHz, while the LISA-TAIJIp becomes more sensitive when reference frequency is lower than 0.9 mHz.

\begin{figure*}[htb]
\includegraphics[width=0.46\textwidth]{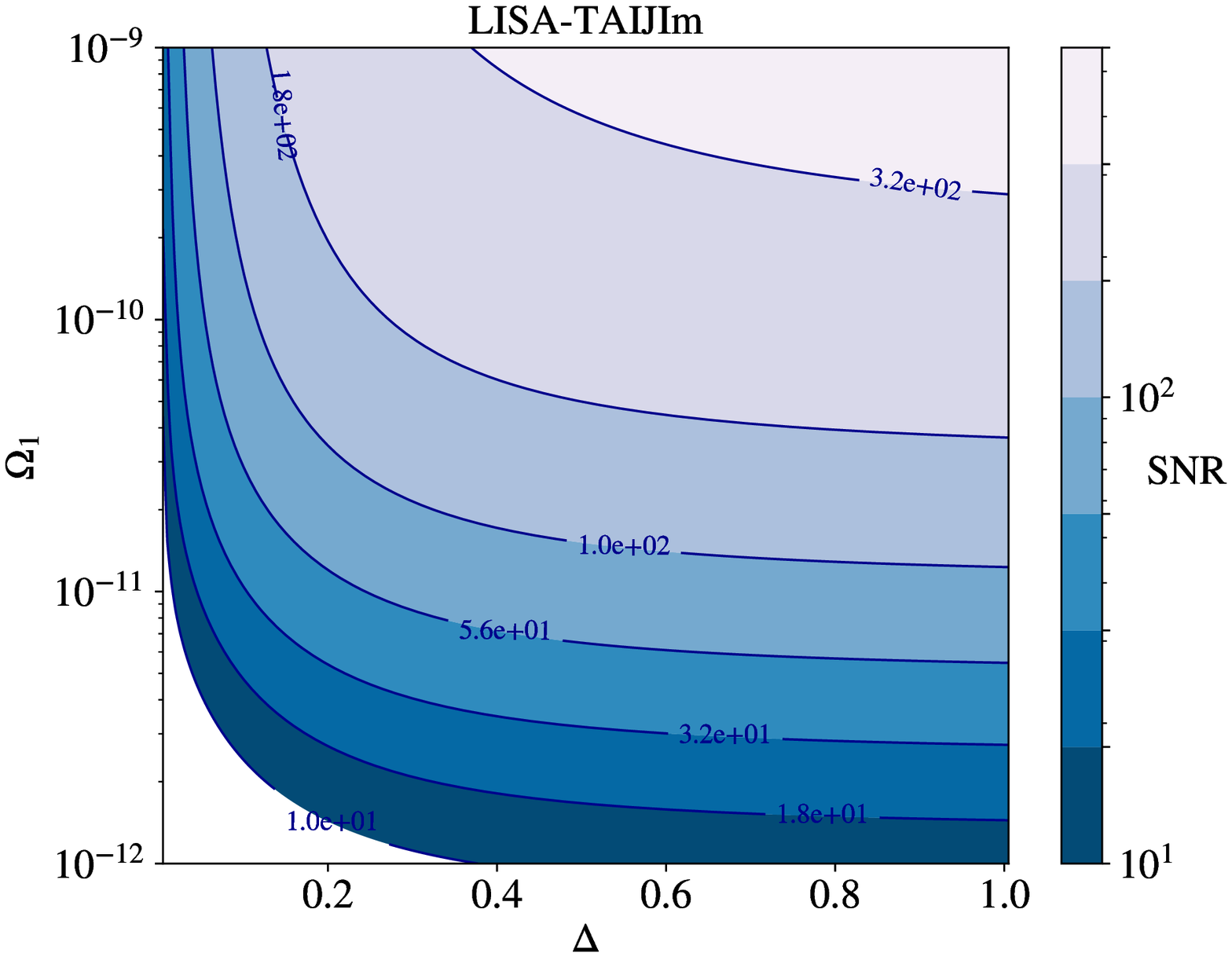} 
\includegraphics[width=0.46\textwidth]{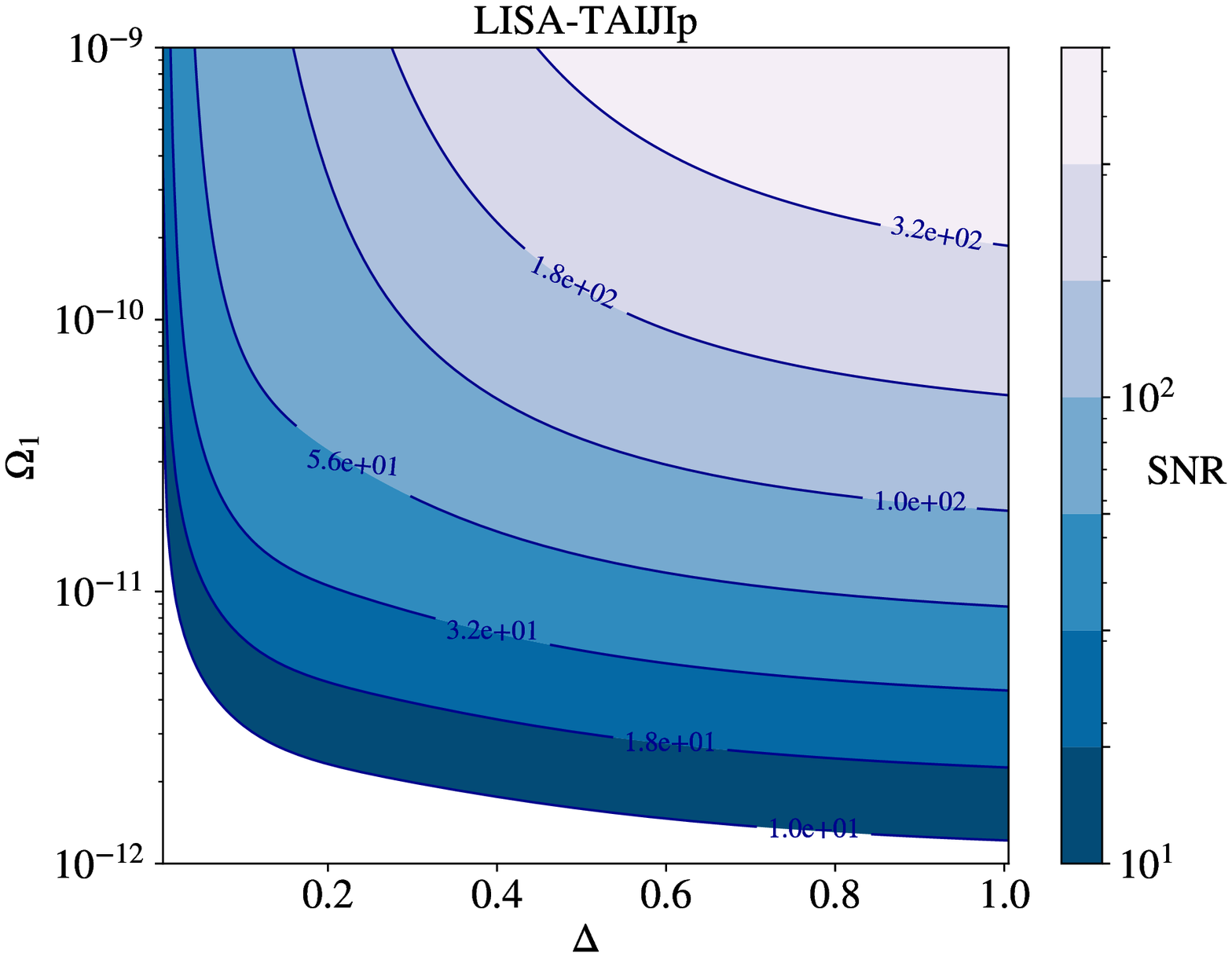} 
\includegraphics[width=0.46\textwidth]{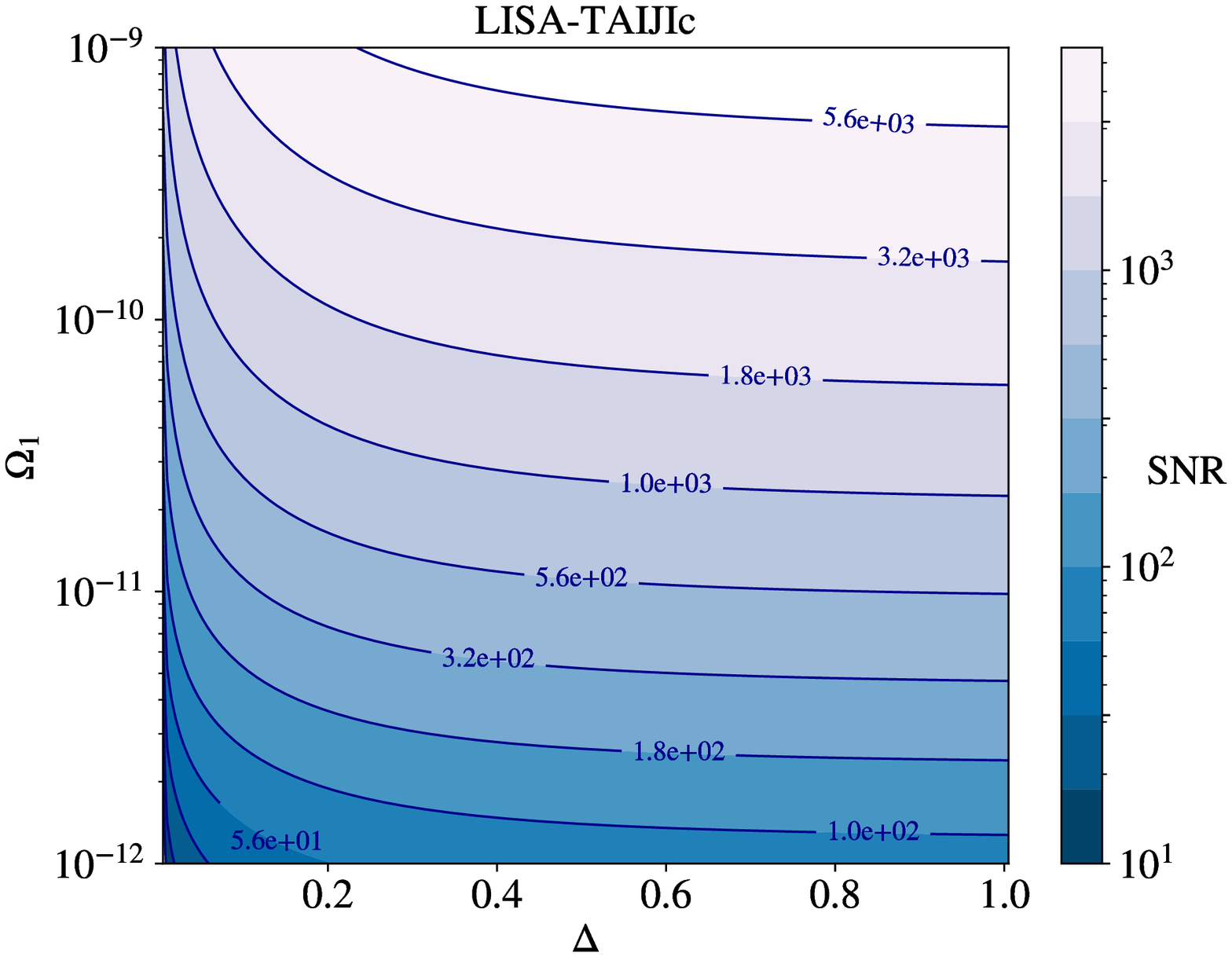} 
\includegraphics[width=0.46\textwidth]{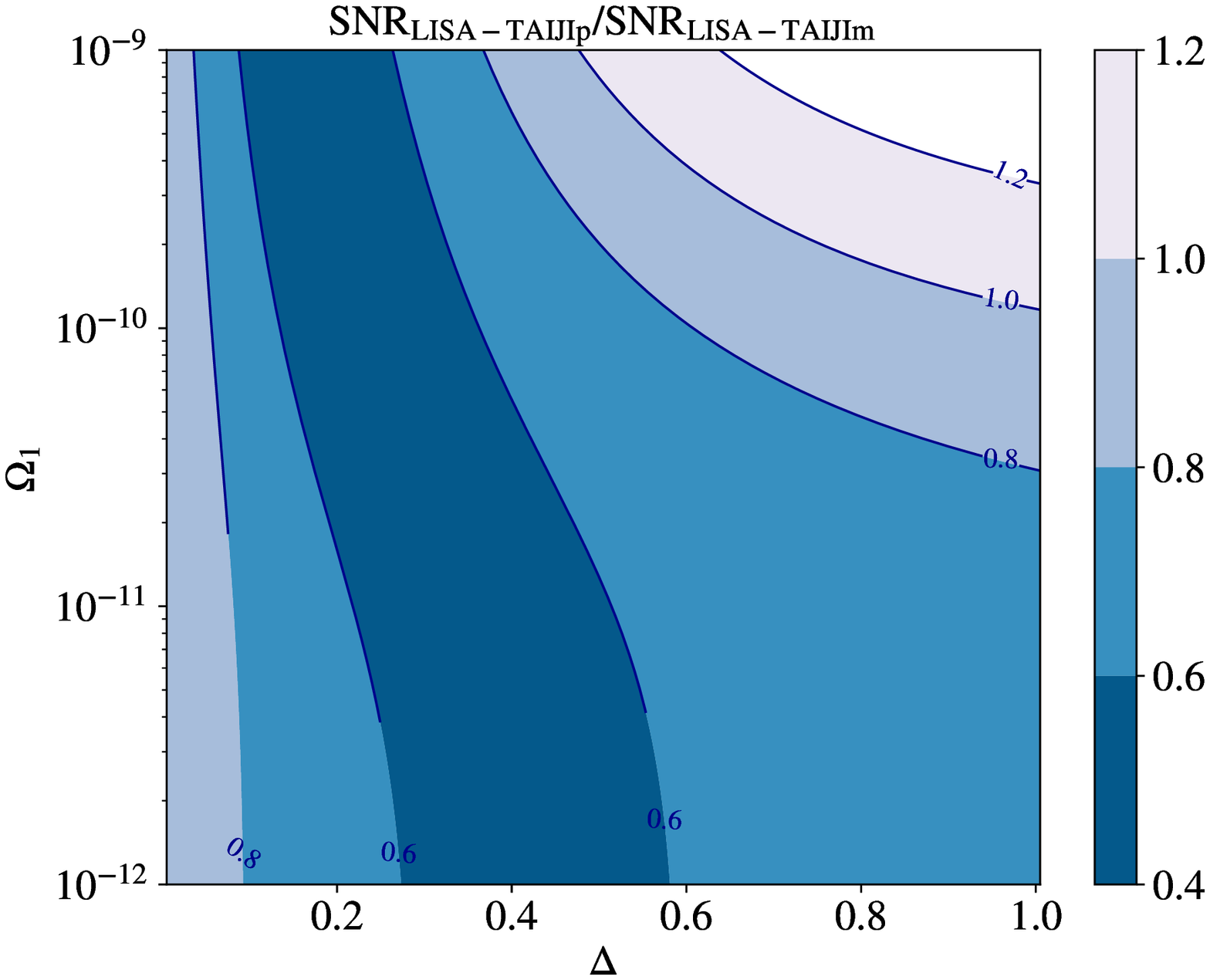}
\caption{The SNR from LISA and TAIJI cross-correlation for the single peaked SGWB signals in parameter space $\Omega_1 \in [10^{-12}, 10^{-9}] \times \Delta \in [0.005, 1]$. The SNR contours for three LISA-TAIJI networks are shown in upper and lower left plots. The SNR ratios of LISA-TAIJIp and LISA-TAIJIm are shown in the lower right plot, the LISA-TAIJIm achieves higher SNR than LISA-TAIJIp configuration for most selected parameters except for higher $\Omega_1$ and $\Delta$ at the upper right corner. \label{fig:SNR_SP} 
}
\end{figure*}

The parameter space ($\Omega_1 \in [10^{-12}, 10^{-9}], \ \Delta \in [0.005, 1]$) in Eq. \eqref{eq:SP_signal} are explored to evaluate the detectability of three networks for single peaked SGWB signal with a fixed reference frequency $f_\mathrm{ref} = 3$ mHz. For all three networks, the larger $\Omega_1$ and larger $\Delta$ values will yield a higher SNR. The SNR from LISA-TAIJIc cross-correlation is much higher than another two networks. The SNR ratios of LISA-TAIJIp and LISA-TAIJIm are shown in the lower right panel of Fig. \ref{fig:SNR_SP}. In the selected parameter space, the LISA-TAIJIp is more sensitive than LISA-TAIJIm only in a small region with both large $\Omega_1$ and large $\Delta$ values, and LISA-TAIJIm could achieve higher SNR than LISA-TAIJIp in most of the parameter space. If the reference frequency $f_\mathrm{ref}$ is the second variable instead of $\Delta$, then we could intuitively expect from Fig. \ref{fig:Omega_sensitivity} that the LISA-TAIJIm would yield higher SNR than LISA-TAIJIp for $f_\mathrm{ref} \succcurlyeq 1$ mHz, and LISA-TAIJIp is more sensitive to single peaked SGWB when reference frequency lower than 1 mHz.   

Comparing the results from LISA-TAIJIm and LISA-TAIJIp for four assumed SGWB signals, the LISA-TAIJIm achieves higher SNRs than LISA-TAIJIp for the most selected parameter spaces, and a decisive factor is that the LISA-TAIJIm has a better sensitivity at the frequencies around 2 mHz which is also the most sensitive band for these two networks. On the other side, the SGWB spectral shapes tend to have stronger signals at a frequency higher than $\sim$1 mHz in selected parameter spaces. Although the LISA-TAIJIp may have a better detectability for other signal assumption(s), the LISA-TAIJIm will be a better configuration for the power-law SGWB inferred from Advanced LIGO and Advanced Virgo runs \cite{LIGOScientific:2019vic,Abbott:2021xxi}.

\section{Discriminating SGWB components by using LISA-TAIJI networks} \label{sec:seperate_components}

As introduced in Section \ref{subsec:SGWB_model}, the SGWB includes astrophysical and cosmological origins. It will be essential to decipher the information in SGWB by discriminating the compositions and retrieving the shapes of the signals.
By using a single LISA mission, \citet{Adams:2010vc,Adams:2013qma} developed a method to separate the stochastic GW background from the instrument noise and Galactic foreground, \citet{Caprini:2019pxz} and \citet{Flauger:2020qyi} demonstrated the reconstruction of the spectral shapes of SGWB, and \citet{Boileau:2020rpg} implemented the SGWB spectral separation by using simulated data from LISA Mock Data Challenge. 
In this section, we will examine the performance of three LISA-TAIJI networks on SGWB components separation. 

\subsection{Fisher matrix analysis}

The Fisher information matrix (FIM) is employed to determine the parameters describing SGWB spectral shapes \cite[and references therein]{Kuroyanagi:2018csn,Smith:2019wny,Boileau:2020rpg,Saffer:2020xsw,Martinovic:2020hru}. Two parts are expected to contribute to the FIM calculation, the first one is from an individual mission's three optimal channels, and the second part is the cross-correlation between two missions.

For LISA or TAIJI mission with six laser links, three orthogonal optimal channels (A, E, and T) could be formed. And the T channel is expected to be a null data stream and could be utilized to characterize the noise of the instruments, and two science data channels, A and E, could be used to detect SGWB signals \cite{Adams:2010vc,Adams:2013qma,Boileau:2020rpg}. In a realistic case, the T channel will not a fully null data stream due to the inequality of arms \cite{Adams:2010vc,Wang:2020a,Wang:1stTDI,Wang:2ndTDI}. And the parameters of SGWB could be inferred from joint three optimal channels \cite{Smith:2019wny,Boileau:2020rpg},
\begin{equation} \label{eq:FIM_self}
\begin{aligned}
 F_{ab, \mathrm{self} } = & \sum_{i = \mathrm{A, E, T} } 2 T_\mathrm{obs} \int^{f_\mathrm{max}}_{0}  \frac{  \frac{\partial S_{h,i}  }{ \partial \theta_a  }  \frac{\partial S_{h,i} }{ \partial \theta_b  } }{ \left[ S_{n, i}  + S_{h, i}  \right]^2 } \mathrm{d} f \\
 = & \sum_{i = \mathrm{A, E, T} } \left(\frac{3 H^2_0}{ 4 \pi ^2} \right)^2 2 T_\mathrm{obs} \int^{f_\mathrm{max}}_{0}  \frac{  \mathcal{R}^2_i \frac{\partial \Omega_{\mathrm{tot},i}  }{ \partial \theta_a  }  \frac{\partial \Omega_{\mathrm{tot},i} }{ \partial \theta_b  } }{ f^6 \left[ S_{n, i}  + S_{h, i} \right]^2 } \mathrm{d} f, 
\end{aligned}
\end{equation}
where $\Omega_\mathrm{tot}$ is energy density of SGWB combining the astrophysical and cosmological components, $\theta_a$ is the SGWB parameter to be determined, and $S_{n, i}$ and $S_{h, i}$ are the PSD of noise and SGWB in TDI $i$ channel, respectively. 

The cross-correlations between two missions are also could used to determine the parameters of SGWB. By assuming the signal is weak compared to the detector noises, the corresponding FIM could be calculated as \cite{Kuroyanagi:2018csn},
\begin{equation} \label{eq:FIM_cross}
\begin{aligned}
 F_{ab, \mathrm{cross} } = & \sum_{i,j = \mathrm{A, E, T} } 2 T_\mathrm{obs} \int^{f_\mathrm{max}}_{0}  \frac{\gamma^2_{ij}  \frac{\partial P_{h}  }{ \partial \theta_a  }  \frac{\partial P_{h} }{ \partial \theta_b  } }{ S^\mathrm{LISA}_{n, i}  S^\mathrm{TAIJI}_{n, j}  } \mathrm{d} f \\ 
= & \sum_{i,j = \mathrm{A, E, T} } \left(\frac{3 H^2_0}{ 4 \pi ^2} \right)^2 2 T_\mathrm{obs} \int^{f_\mathrm{max}}_{0}  \frac{  \gamma^2_{ij}  \frac{\partial \Omega_\mathrm{tot}  }{ \partial \theta_a  }  \frac{\partial \Omega_\mathrm{tot} }{ \partial \theta_b  } }{ f^6 S^\mathrm{LISA}_{n, i}  S^\mathrm{TAIJI}_{n, j} } \mathrm{d} f, 
\end{aligned}
\end{equation}
where $\gamma_{ij}$ is the overlap reduction function between LISA's $i$ channel and TAIJI's $j$ channel as calculated in Eq. \eqref{eq:gamma_ij}, $P_h$ is the PSD of SGWB defined in Eq. \eqref{eq:P_h}.
The FIM of the joint LISA-TAIJI network is obtained by summing up the FIM from two missions' cross-correlation and two individual missions,
\begin{equation}
F_{ab, \mathrm{joint} } =  F_{ab, \mathrm{cross} } + \sum^{\rm TAIJI}_{m = \rm LISA } F^m_{ab, \mathrm{self}} \label{eq:FIM_tot}
\end{equation}
The variance-covariance matrix of the parameters will be
\begin{equation}
\begin{aligned}
 \left\langle \delta \theta_i \delta \theta_j  \right\rangle = \left( F^{-1}_{ab, \mathrm{joint} } \right)_{ij} + \mathcal{O}({\rho}^{-1})
\overset{{\rho} \gg 1}{\simeq } \left( F^{-1}_{ab, \mathrm{joint} } \right)_{ij} . \\
\end{aligned}
\end{equation}
The standard deviations $\sigma_i$ of the parameter $i$ is
\begin{equation}
\begin{aligned}
\sigma_{i} & \overset{{\rho} \gg 1}{\simeq} \sqrt{ \left( F^{-1}_{ab, \mathrm{joint} } \right)_{ii}  }. \\
\end{aligned}
\end{equation}

\subsection{Determining SGWB parameters}

The total SGWB signal is composed of astrophysical and cosmological parts, $\Omega_\mathrm{tot} = \Omega_\mathrm{astro}  + \Omega_\mathrm{cosmos} $. The shape of the astrophysical component is assumed to be a power-law shape in Eq. \eqref{eq:PL_signal}, and the cosmological part is represented by flat, broken power-law, or single peaked shapes as specified in Eqs. \eqref{eq:flat_signal}-\eqref{eq:SP_signal}. Therefore, three combinations, ($\Omega_\mathrm{PL}+ \Omega_\mathrm{flat}$, $\Omega_\mathrm{PL}+ \Omega_\mathrm{BPL}$, $\Omega_\mathrm{PL}+ \Omega_\mathrm{SP}$), are constructed for SGWB observed by LISA-TAIJI networks, and the constraints on parameters describing the spectral shapes are investigated. And the fiducial values are utilized to characterize the shapes of each selected SGWB signal.

For the power-law signal, the parameters, $\Omega_0$ and $\alpha_0$ in Eq. \eqref{eq:PL_signal} are selected to be determined, and the partial derivatives of SGWB with respect to these two parameters are
\begin{align}
\frac{\partial \Omega_\mathrm{PL} }{ \partial \log_{10} \Omega_{0}  }  = & \Omega_\mathrm{PL} \ln 10, \\
\frac{\partial \Omega_\mathrm{PL} }{ \partial \alpha_0 }  = & \Omega_\mathrm{PL} \ln \frac{f}{ 1 \ \mathrm{mHz} }.
\end{align}
When the flat is assumed to be the cosmological SGWB shape, the partial derivative of $\Omega_\mathrm{tot} = \Omega_\mathrm{PL} + \Omega_\mathrm{flat} $ to the amplitude of energy density $\Omega_1$ will be, 
\begin{align}
\frac{\partial \Omega_\mathrm{tot} }{ \partial \log_{10} \Omega_{1}  }  = &  \Omega_\mathrm{flat} \ln 10. 
\end{align}
When SGWB is described by the broken power-law model, three parameters ($\Omega_1$, $\alpha_1$, and $\alpha_2$) are selected to be determined, and the partial derivatives of $\Omega_\mathrm{tot} = \Omega_\mathrm{PL} + \Omega_\mathrm{BPL}$ to three parameters will be  
\begin{align}
\frac{\partial \Omega_\mathrm{tot} }{ \partial \log_{10} \Omega_{1}  }  = &  \Omega_\mathrm{BPL} \ln 10,  \\
\frac{\partial \Omega_\mathrm{tot} }{ \partial \alpha_1 }  = & \Omega_\mathrm{BPL} \left[ \ln \frac{f}{ 10 \ \mathrm{mHz} }  \right. \\ \notag
& \left. -  \frac{1}{\Delta} \ln \left( 1 + 0.75 \left( \frac{f}{ 10 \ \mathrm{mHz} } \right)^\Delta \right) \right], \\
\frac{\partial \Omega_\mathrm{tot} }{ \partial \alpha_2 }  = & \Omega_\mathrm{BPL} \frac{1}{\Delta} \ln \left( 1 + 0.75 \left( \frac{f}{ 10 \ \mathrm{mHz} } \right)^\Delta \right).
\end{align}
If the cosmological SGWB is a single peaked shape, $\Omega_\mathrm{tot} = \Omega_\mathrm{PL} + \Omega_\mathrm{SP}$, the derivatives of $\Omega_\mathrm{tot}$ to $\Omega_1$ and $\Delta$ are
\begin{align}
\frac{\partial \Omega_\mathrm{tot} }{ \partial \log_{10} \Omega_{1}  }  = &  \Omega_\mathrm{SP} \ln 10, \\
\frac{\partial \Omega_\mathrm{tot} }{ \partial \Delta }  = &  \Omega_\mathrm{SP}  \frac{ 2 \left[ \log_{10} ( f / 3 \ \mathrm{mHz}  ) \right]^2}{ \Delta^3 }.
\end{align}

\begin{figure}[hbt]
\includegraphics[width=0.48\textwidth]{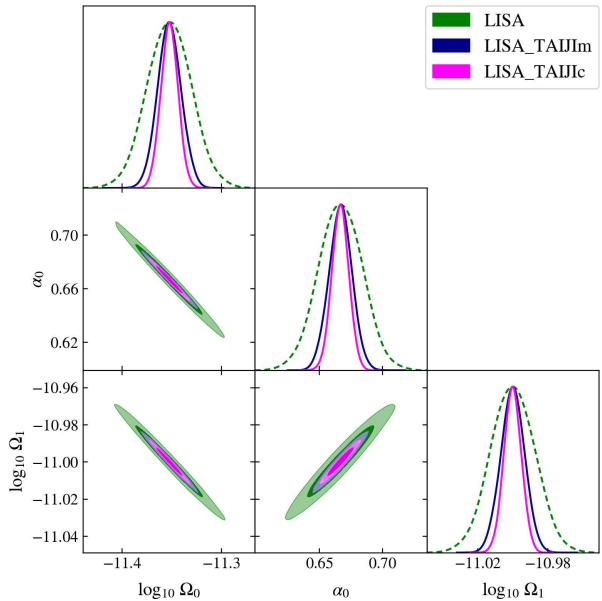} 
\caption{The parameter corner plot of SGWB signal model $\Omega_\mathrm{tot} = \Omega_\mathrm{PL} + \Omega_\mathrm{flat}$ with single LISA (green), LISA-TAIJIm (blue) and LISA-TAIJIc (magenta) configurations. The uncertainties of parameters from LISA and LISA-TAIJI networks are shown in Table \ref{tab:PL_flat}. (The result from LISA-TAIJIp is not shown in the plot because it is highly overlapped with LISA-TAIJIm result).  \label{fig:corner_flat} 
}
\end{figure}

For each SGWB combination, the FIM from a LISA-TAIJI network is obtained by implementing Eq. \eqref{eq:FIM_tot}. And the distribution of parameters could be made by 
\begin{equation}
p ( \Delta \vec{\theta} ) = \sqrt{ \det \left( \frac{F_{ab}}{ 2 \pi} \right)} \exp \left( - \frac{1}{2} \delta \vec{\theta}^{T} F_{ab} \delta \vec{\theta} \right).
\end{equation}
The corner plot for the $\Omega_\mathrm{tot} = \Omega_\mathrm{PL} + \Omega_\mathrm{flat} $ case is shown in Fig. \ref{fig:corner_flat}, and the uncertainties of three parameters from different mission configurations are listed in Table \ref{tab:PL_flat}. As Table \ref{tab:PL_flat} showed, the constraints on parameters from LISA-TAIJIm and LISA-TAIJIp are comparable and better than the LISA mission by a factor of $\sim$1.8. The LISA-TAIJIc could further narrow down uncertainties by a factor of $\sim$2.5 compared to LISA.

\begin{table}[tbh]
\caption{\label{tab:PL_flat} The $1\sigma$ uncertainties of parameters of SGWB model $\Omega_\mathrm{tot} = \Omega_\mathrm{PL} + \Omega_\mathrm{flat}$ from LISA and LISA-TAIJI configurations.
}
\begin{ruledtabular}
\begin{tabular}{ccccc}
Parameter & LISA & LISA-TAIJIm &  LISA-TAIJIp  & LISA-TAIJIc  \\
\hline
$\delta \log_{10} \Omega_0$  & 2.24e-2 & 1.18e-2 & 1.16e-2 & 8.65e-3 \\
$\delta \alpha_0$ & 1.74e-2 & 9.04e-3 & 8.90e-3  & 6.71e-3  \\
$\delta \log_{10} \Omega_1$ & 1.27e-2 & 6.77e-3 & 6.64e-3 & 4.99e-3 \\
\end{tabular}
\end{ruledtabular}
\end{table}

\begin{figure*}[htb]
\includegraphics[width=0.85\textwidth]{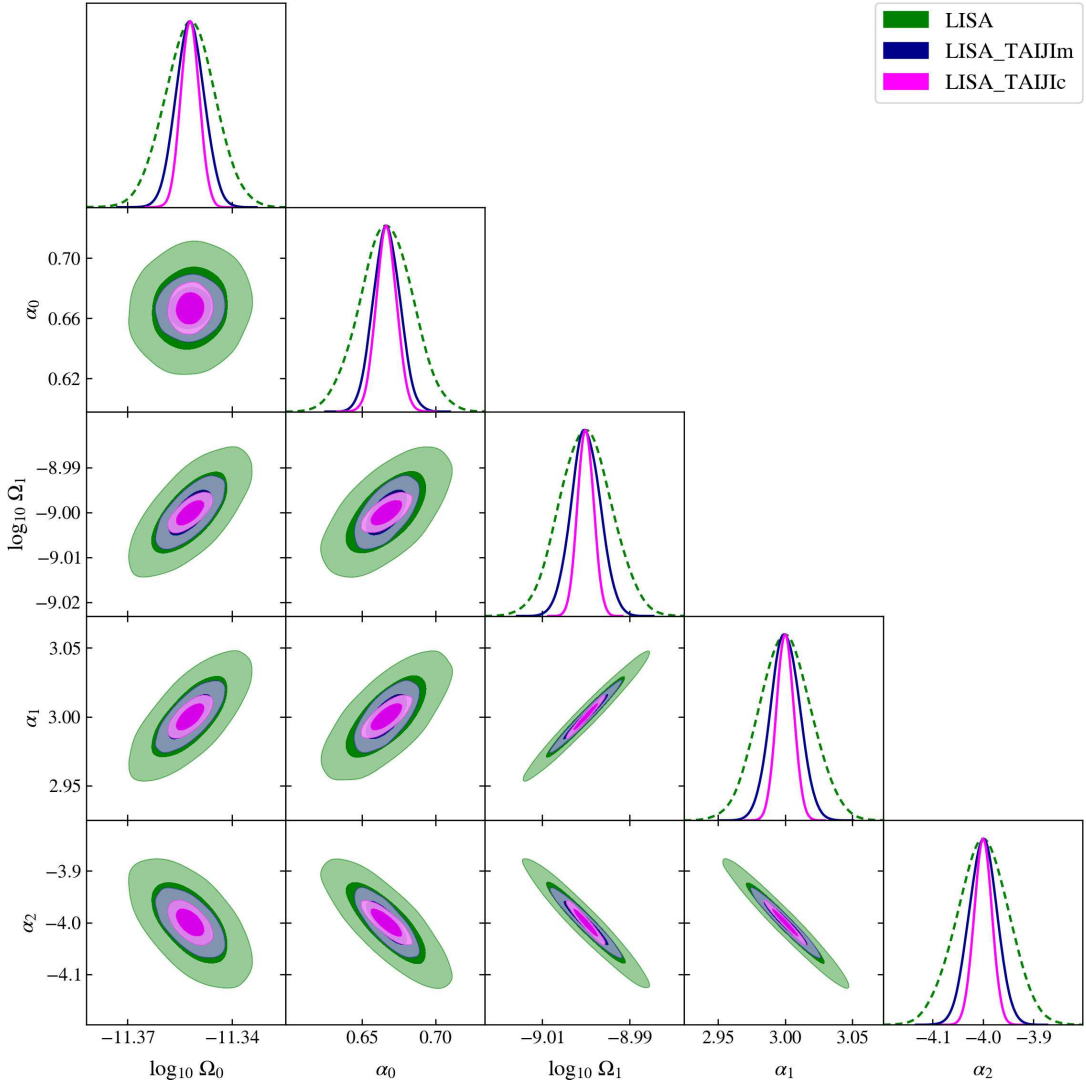} 
\caption{The parameter corner plot of SGWB signal model $\Omega_\mathrm{tot} = \Omega_\mathrm{PL} + \Omega_\mathrm{BPL}$ with single LISA (green), LISA-TAIJIm (blue) and LISA-TAIJIc (magenta) configurations. The uncertainties of parameters from LISA and LISA-TAIJI networks are shown in Table \ref{tab:PL_BPL}. (The result from LISA-TAIJIp is not shown in the plot because it is highly overlapped with LISA-TAIJIm result). \label{fig:corner_BPL} 
}
\end{figure*}

The corner plot and uncertainties of parameters for $\Omega_\mathrm{tot} = \Omega_\mathrm{PL} + \Omega_\mathrm{BPL} $ scenario are shown in Fig. \ref{fig:corner_BPL} and Table \ref{tab:PL_BPL}, respectively.
Compared to PL+flat results in Table \ref{tab:PL_flat}, the uncertainties of the amplitude of the energy density $\Omega_0$ and $\Omega_1$ could be narrowed in the PL+BPL case which means the PL and BPL components could be better separated. For performance between mission configurations, the capabilities of the LISA-TAIJIm and LISA-TAIJIp are still comparable and could determine the parameters with a better accuracy than LISA by a factor of $\sim$1.8. The parameter resolutions from the LISA-TAIJIc network could be $\sim$2.5-3.2 times better than LISA for the selected five parameters. 

\begin{table}[tbh]
\caption{\label{tab:PL_BPL} The FIM uncertainties of parameters of SGWB model $\Omega_\mathrm{tot} = \Omega_\mathrm{PL} + \Omega_\mathrm{BPL}$ from LISA and LISA-TAIJI networks.
}
\begin{ruledtabular}
\begin{tabular}{ccccc}
Parameter & LISA & LISA-TAIJIm &  LISA-TAIJIp  & LISA-TAIJIc  \\
\hline
$\delta \log_{10} \Omega_0$  & 7.24e-3 &  4.13e-3 & 4.20e-3  & 2.60e-3  \\
$\delta \alpha_0$ & 1.78e-2  & 9.07e-3  & 9.08e-3  & 7.08e-3 \\
$\delta \log_{10} \Omega_1$ & 5.84e-3  & 3.36e-3  & 3.38e-3  & 1.83e-3 \\
$\delta \alpha_1$ &  1.93e-2 & 1.08e-2  & 1.09e-2  & 6.40e-3  \\
$\delta \alpha_2$ & 5.04e-2  & 2.79e-2  &  2.81e-2  & 1.76e-2 \\
\end{tabular}
\end{ruledtabular}
\end{table}

\begin{figure*}[htb]
\includegraphics[width=0.75\textwidth]{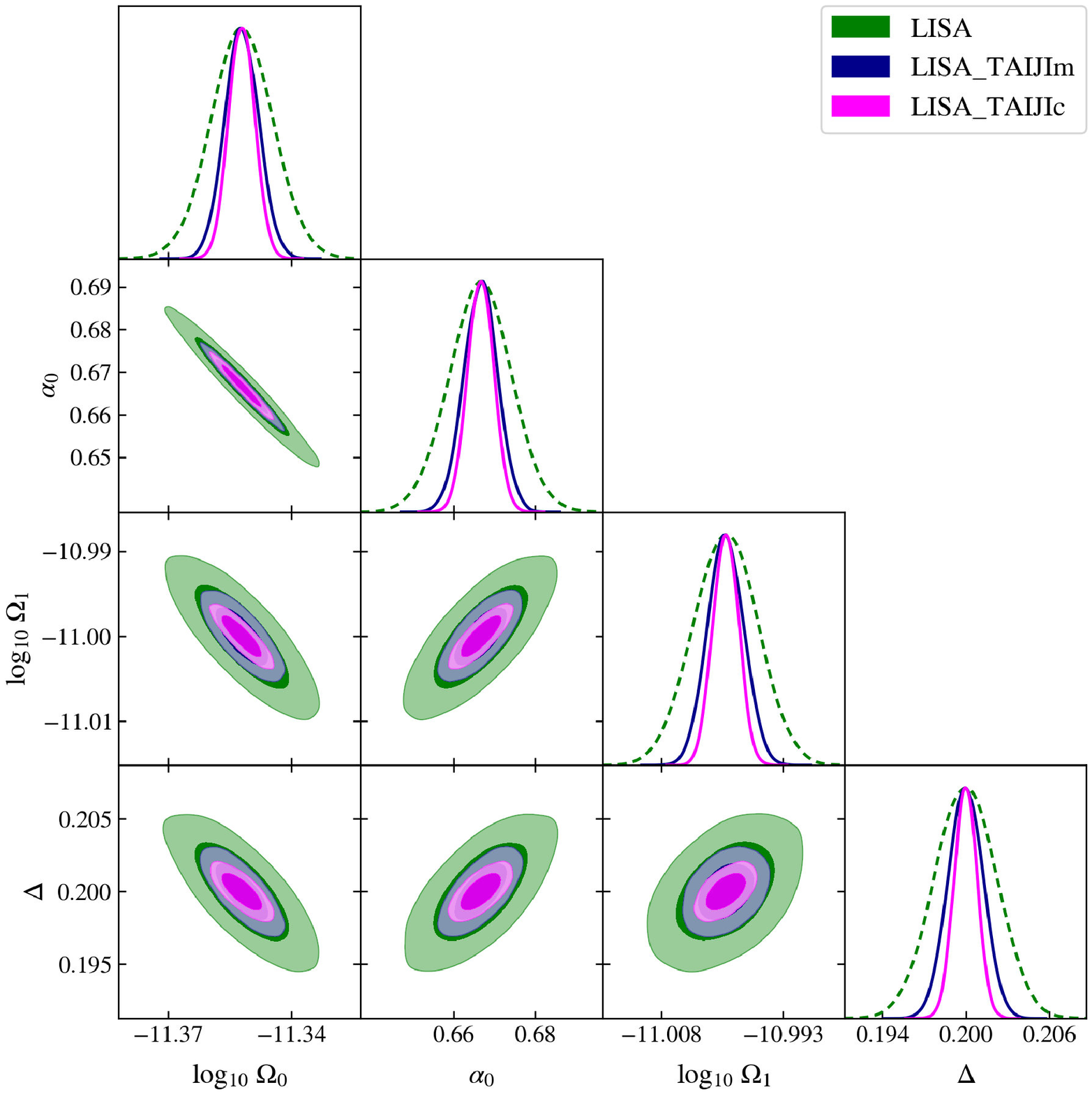} 
\caption{The parameter corner plot of SGWB signal model $\Omega_\mathrm{tot} = \Omega_\mathrm{PL} + \Omega_\mathrm{SP}$ with single LISA (green), LISA-TAIJIm (blue) and LISA-TAIJIc (magenta) configurations. The uncertainties of parameters from LISA and LISA-TAIJI networks are shown in Table \ref{tab:PL_SP}. (The result from LISA-TAIJIp is not shown in the plot because it is highly overlapped with LISA-TAIJIm result).  \label{fig:corner_single_peak} 
}
\end{figure*}

Fig. \ref{fig:corner_single_peak} and Table \ref{tab:PL_SP} show the parameter constraints from LISA and LISA-TAIJI networks for SGWB signals $\Omega_\mathrm{tot} = \Omega_\mathrm{PL} + \Omega_\mathrm{SP}$. Similar to the previous two scenarios, both LISA-TAIJIm and LISA-TAIJIp networks could reduce parameter uncertainties by a factor of $\sim$1.8 compared to LISA, and joint observation from LISA and TAIJIc could promote the factors to $\sim$2.3-2.6 for the selected parameters.  
 
\begin{table}[tbh]
\caption{\label{tab:PL_SP} The $1\sigma$ uncertainties of the parameters of SGWB model $\Omega_\mathrm{tot} = \Omega_\mathrm{PL} + \Omega_\mathrm{SP}$ from LISA and LISA-TAIJIc networks.
}
\begin{ruledtabular}
\begin{tabular}{ccccc}
Parameter & LISA & LISA-TAIJIm &  LISA-TAIJIp  & LISA-TAIJIc  \\
\hline
$\delta \log_{10} \Omega_0$  & 7.60e-3 & 4.10e-3 & 4.08e-3 & 3.20e-3 \\
$\delta \alpha_0$ & 7.75e-3 & 4.15e-3 & 4.13e-3  & 3.28e-3  \\
$\delta \log_{10} \Omega_1$ & 4.00e-3 & 2.21e-3 & 2.22e-3 & 1.58e-3 \\
$\delta \Delta$ & 2.24e-3 & 1.26e-3 & 1.27e-3  & 8.47e-4 \\
\end{tabular}
\end{ruledtabular}
\end{table}

As we can deduce from these results, for a stochastic signal, LISA-TAIJIm or LISA-TAIJIp joint observation could determine the parameters with a better accuracy than LISA by a factor of $\sim$1.8. And this improvement is largely contributed by the two missions' individual observations which correspond to the term $F_{ab, \mathrm{self}}$ in Eq. \eqref{eq:FIM_tot}. The reason is that comparing the curves in Fig. \ref{fig:Omega_sensitivity}, the sensitivity for LISA or TAIJI obtained from Eq. \eqref{eq:Omega_mission_sensitivity} is much better than the sensitivities for LISA-TAIJIm or LISA-TAIJIp cross-correlation from Eq. \eqref{eq:Omega_LISA_TAIJI_sensitivty}. Therefore, we can expect the FIM from $F_{ab, \mathrm{self}}$ (from Eq. \eqref{eq:FIM_self}) should contribute more constraints on parameter measurements than $F_{ab, \mathrm{cross}}$ (from Eq. \eqref{eq:FIM_cross}). For the LISA-TAIJIc network, the FIM from cross-correlation could be comparable to a single LISA mission. However, a caveat is that a weak signal approximation is applied by using Eq. \eqref{eq:FIM_cross}, and the FIM may overestimate the contribution of cross-correlation from two detectors for a strong SGWB signal.

\section{Conclusions} \label{sec:conclusions}

In this work, we investigate the detectability of three LISA-TAIJI networks to the isotropic SGWB. For the colocated and coplanar LISA-TAIJIc network, it can detect the SGWB with an optimal sensitivity. The TAIJIp and TAIJIm are placed at locations $1 \times 10^8$ km from the LISA detector, compared to the LISA-TAIJIm network, the LISA and TAIJIp will have stronger correlated data streams, and their cross-correlation will be more sensitive to SGWB at frequencies lower than $\sim$1 mHz. However, the LISA-TAIJIm has a better sensitivity than LISA-TAIJIp in the frequency band $\sim$[1, 8] mHz which will be decisive to promote its detectability to SGWB signals. In the selected parameter spaces for four assumed SGWB shapes, the performance from LISA-TAIJIm is competitive with LISA-TAIJIp configuration or even better than LISA-TAIJIp for selected fiducial cases.

The capabilities of three LISA-TAIJI networks to determine parameters and separate the SGWB components are also examined. The Fisher matrix algorithm for weak signal approximation is employed to perform this investigation. By combining two kinds of SGWB signals from assumed fiducial cases, the LISA-TAIJIc could resolve the parameters of SGWB signals with better accuracy than LISA by a factor of $\sim$2.3-3.2. For LISA-TAIJIm and LISA-TAIJIp networks, both of them can determine parameters with a $\sim$1.8 times better resolution than LISA. The higher parameter resolutions would be used to more precisely reconstruct the shapes of SGWB and discriminate the components. The joint observation is also expected to better discern Galactic foreground and instrument noises from SGWB, and we commit it as our next work. 

Considering the previous investigations in \cite{Wang:alternative}, although LISA-TAIJIc is optimal for the SGWB detection, it would not be an optimal choice to observe the MBH binaries compared to the large separated LISA-TAIJIm or LISA-TAIJIp networks. Moreover, the colocated detectors may be subject to the same space environments and cause correlated noises. For the selected SGWB models in this work, the detectability of LISA-TAIJIm network to the SGWB is competitive with LISA-TAIJIp. And comparable performances from these two networks could be also expected for other SGWB models. In addition, the LISA-TAIJIm could achieve a better detectability to the MBH binaries than the LISA-TAIJIp configuration. Therefore, the LISA-TAIJIm network could be recognized as an optimal configuration to fulfill the joint observation.

\begin{acknowledgments}
This work was supported by NSFC Nos. 12003059 and 11773059, the Strategic Priority Research Program of the Chinese Academy of Sciences under Grants No. XDA15021102. This work made use of the High Performance Computing Resource in the Core Facility for Advanced Research Computing at Shanghai Astronomical Observatory.
The calculations in this work are performed by using the python packages $\mathsf{numpy}$ \cite{harris2020array} and $\mathsf{scipy}$ \cite{2020SciPy-NMeth}, and the plots are make by utilizing $\mathsf{matplotlib}$ \cite{Hunter:2007ouj} and $\mathsf{GetDist}$ \cite{Lewis:2019xzd}.
\end{acknowledgments}

\appendix

\section{Response formulation of laser link to GW} \label{sec:appendix_response}

For a source locating at ecliptic longitude $\lambda$ and latitude $\theta$ (in the solar-system barycentric coordinates), the GW propagation vector will be
\begin{equation} \label{eq:source_vec}
 \hat{k}  = -( \cos \lambda \cos \theta, \sin \lambda \cos \theta ,  \sin \theta ).
\end{equation}
The $+$, $\times$ polarization tensors of the GW signal combining source's inclination angle $\iota$ are
\begin{equation} \label{eq:polarizations-response}
\begin{aligned}
{\rm e}_{+} & \equiv \mathcal{O}_1 \cdot
\begin{pmatrix}
1 & 0 & 0 \\
0 & -1 & 0 \\
0 & 0 & 0
\end{pmatrix}
\cdot \mathcal{O}^T_1 \times \frac{1+\cos^2 \iota}{2} ,
\\
{\rm e}_{\times} &  \equiv \mathcal{O}_1 \cdot
\begin{pmatrix}
0 & 1 & 0\\
1 & 0 & 0 \\
0 & 0 & 0
\end{pmatrix}
\cdot \mathcal{O}^T_1 \times i (- \cos \iota ),
\end{aligned}
\end{equation}
with
\begin{widetext}
\begin{equation}
\mathcal{O}_1 =
\begin{pmatrix}
\sin \lambda \cos \psi - \cos \lambda \sin \theta \sin \psi & -\sin \lambda \sin \psi - \cos \lambda \sin \theta \cos \psi & -\cos \lambda \cos \theta  \\
     -\cos \lambda \cos \psi - \sin \lambda \sin \theta \sin \psi & \cos \lambda \sin \psi - \sin \lambda \sin \theta \cos \psi & -\sin \lambda \cos \theta  \\
         \cos \theta \sin \psi & \cos \theta \cos \psi & -\sin \theta
\end{pmatrix},
\end{equation}
where $\psi$ is polarization angle. The response to the GW in laser link from S/C$i$ to $j$ will be
\begin{equation} \label{eq:y_ij}
\begin{aligned}
y^{h}_{ij} (f) =&  \frac{ \sum_\mathrm{p} \hat{n}_{ij} \cdot {\mathrm{ e_p}} \cdot \hat{n}_{ij} }{2 (1 - \hat{n}_{ij} \cdot \hat{k} ) }
 \times \left[  \exp( 2 \pi i f (L_{ij} + \hat{k} \cdot p_i ) ) -  \exp( 2 \pi i f  \hat{k} \cdot p_j )  \right] ,
\end{aligned}
\end{equation}
\end{widetext}
where $\hat{n}_{ij}$ is the unit vector from S/C$i$ to $j$, $L_{ij}$ is the arm length from S/C$i$ to $j$, $p_i$ is the position of the S/C$i$ in the solar-system barycentric ecliptic coordinates. The response of Michelson-X channel to GW will be 
\begin{equation} \label{eq:TDI_Fh}
\begin{aligned}
F^h_{ \rm X} (f) =& (-\Delta_{21} + \Delta_{21}  \Delta_{13}  \Delta_{31})  y^h_{12} \\
        & + (-1 + \Delta_{13}  \Delta_{31} )  y^h_{21} \\
        & + (\Delta_{31} - \Delta_{31}  \Delta_{12}  \Delta_{21})  y^h_{13} \\
        & + ( 1 - \Delta_{12}  \Delta_{21} )  y^h_{31},
\end{aligned}
\end{equation}
where $\Delta_{ij} = \exp(2 \pi i f L_{ij})$.

\section{Overlap reduction function}

The \textit{overlap reduction function} is introduced to indicate the cross-correlation between two laser interferometers \cite{Christensen:1992wi}.
The overlap reduction function for three LISA-TAIJI networks have been calculated in \cite{Wang:alternative}, the colocated and coplanar LISA-TAIJIc network yields unity overlap reduction function for frequency lower than 10 mHz which should be an optimal configuration for the SGWB observation. The overlap reduction function change the sign between two detectors' characteristic frequencies gap [$\frac{c}{2 L_\mathrm{TAIJI}} = 50$ mHz, $\frac{c}{2 L_\mathrm{LISA}} = 60$ mHz] ($c$ is the speed of the light). The most misaligned LISA-TAIJIm yield the worst overlap function in the three networks for frequencies lower than a critical frequency  $f_\mathrm{crit} \simeq c/(2 D_\mathrm{sep}) \simeq  1.5 $ mHz considering the separation, $ D_\mathrm{sep} = 1 \times 10^8$ km, between LISA and TAIJIp/TAIJIm \cite{Romano:2016dpx}. When the frequency is higher than the critical frequency, the overlap reduction functions of both LISA-TAIJIm and LISA-TAIJIp oscillate and quickly approach zero. 

\begin{figure}
\includegraphics[width=0.45\textwidth]{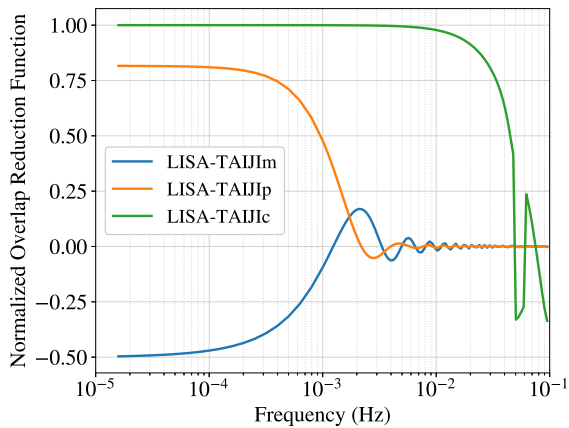} 
\caption{The normalized overlap reduction functions of three LISA-TAIJI networks tensor polarization. The overlap reduction functions of LISA-TAIJI network are from previous work \cite{Wang:alternative}. Each curve represents the sum of the optimal channels from two missions, $\gamma_\mathrm{sum} = \sum_{i,j = \mathrm{A, E, T}} \gamma_{ij} $, and the normalization is implemented to make $\gamma_\mathrm{sum}=1$ when two detectors are coplanar and colocated. The LISA-TAIJIc network has the strongest cross-correlation with each other, and the LISA-TAIJIm network yields the worst cross-correlation in the three configurations for the isotropic SGWB observation. \label{fig:overlap_reduction_fn}
}
\end{figure}

\nocite{*}
\bibliography{apsref}

\providecommand{\noopsort}[1]{}\providecommand{\singleletter}[1]{#1}%
\begin{thebibliography}{66}%
\makeatletter
\providecommand \@ifxundefined [1]{%
 \@ifx{#1\undefined}
}%
\providecommand \@ifnum [1]{%
 \ifnum #1\expandafter \@firstoftwo
 \else \expandafter \@secondoftwo
 \fi
}%
\providecommand \@ifx [1]{%
 \ifx #1\expandafter \@firstoftwo
 \else \expandafter \@secondoftwo
 \fi
}%
\providecommand \natexlab [1]{#1}%
\providecommand \enquote  [1]{``#1''}%
\providecommand \bibnamefont  [1]{#1}%
\providecommand \bibfnamefont [1]{#1}%
\providecommand \citenamefont [1]{#1}%
\providecommand \href@noop [0]{\@secondoftwo}%
\providecommand \href [0]{\begingroup \@sanitize@url \@href}%
\providecommand \@href[1]{\@@startlink{#1}\@@href}%
\providecommand \@@href[1]{\endgroup#1\@@endlink}%
\providecommand \@sanitize@url [0]{\catcode `\\12\catcode `\$12\catcode
  `\&12\catcode `\#12\catcode `\^12\catcode `\_12\catcode `\%12\relax}%
\providecommand \@@startlink[1]{}%
\providecommand \@@endlink[0]{}%
\providecommand \url  [0]{\begingroup\@sanitize@url \@url }%
\providecommand \@url [1]{\endgroup\@href {#1}{\urlprefix }}%
\providecommand \urlprefix  [0]{URL }%
\providecommand \Eprint [0]{\href }%
\providecommand \doibase [0]{https://doi.org/}%
\providecommand \selectlanguage [0]{\@gobble}%
\providecommand \bibinfo  [0]{\@secondoftwo}%
\providecommand \bibfield  [0]{\@secondoftwo}%
\providecommand \translation [1]{[#1]}%
\providecommand \BibitemOpen [0]{}%
\providecommand \bibitemStop [0]{}%
\providecommand \bibitemNoStop [0]{.\EOS\space}%
\providecommand \EOS [0]{\spacefactor3000\relax}%
\providecommand \BibitemShut  [1]{\csname bibitem#1\endcsname}%
\let\auto@bib@innerbib\@empty
\bibitem [{\citenamefont {Wang}\ \emph
  {et~al.}(2021{\natexlab{a}})\citenamefont {Wang}, \citenamefont {Ni},
  \citenamefont {Han}, \citenamefont {Xu},\ and\ \citenamefont
  {Luo}}]{Wang:alternative}%
  \BibitemOpen
  \bibfield  {author} {\bibinfo {author} {\bibfnamefont {G.}~\bibnamefont
  {Wang}}, \bibinfo {author} {\bibfnamefont {W.-T.}\ \bibnamefont {Ni}},
  \bibinfo {author} {\bibfnamefont {W.-B.}\ \bibnamefont {Han}}, \bibinfo
  {author} {\bibfnamefont {P.}~\bibnamefont {Xu}},\ and\ \bibinfo {author}
  {\bibfnamefont {Z.}~\bibnamefont {Luo}},\ }\bibfield  {title} {\bibinfo
  {title} {{Alternative LISA-TAIJI networks}},\ }\href
  {https://doi.org/10.1103/PhysRevD.104.024012} {\bibfield  {journal} {\bibinfo
   {journal} {Phys. Rev. D}\ }\textbf {\bibinfo {volume} {104}},\ \bibinfo
  {pages} {024012} (\bibinfo {year} {2021}{\natexlab{a}})},\ \Eprint
  {https://arxiv.org/abs/2105.00746} {arXiv:2105.00746 [gr-qc]} \BibitemShut
  {NoStop}%
\bibitem [{\citenamefont {Abbott}\ \emph
  {et~al.}(2016{\natexlab{a}})\citenamefont {Abbott} \emph
  {et~al.}}]{Abbott:2016blz}%
  \BibitemOpen
  \bibfield  {author} {\bibinfo {author} {\bibfnamefont {B.~P.}\ \bibnamefont
  {Abbott}} \emph {et~al.} (\bibinfo {collaboration} {LIGO Scientific,
  Virgo}),\ }\bibfield  {title} {\bibinfo {title} {{Observation of
  Gravitational Waves from a Binary Black Hole Merger}},\ }\href
  {https://doi.org/10.1103/PhysRevLett.116.061102} {\bibfield  {journal}
  {\bibinfo  {journal} {Phys. Rev. Lett.}\ }\textbf {\bibinfo {volume} {116}},\
  \bibinfo {pages} {061102} (\bibinfo {year} {2016}{\natexlab{a}})},\ \Eprint
  {https://arxiv.org/abs/1602.03837} {arXiv:1602.03837 [gr-qc]} \BibitemShut
  {NoStop}%
\bibitem [{\citenamefont {Abbott}\ \emph
  {et~al.}(2017{\natexlab{a}})\citenamefont {Abbott} \emph
  {et~al.}}]{GW170817:detection}%
  \BibitemOpen
  \bibfield  {author} {\bibinfo {author} {\bibfnamefont {B.~P.}\ \bibnamefont
  {Abbott}} \emph {et~al.} (\bibinfo {collaboration} {LIGO Scientific,
  Virgo}),\ }\bibfield  {title} {\bibinfo {title} {{GW170817: Observation of
  Gravitational Waves from a Binary Neutron Star Inspiral}},\ }\href
  {https://doi.org/10.1103/PhysRevLett.119.161101} {\bibfield  {journal}
  {\bibinfo  {journal} {Phys. Rev. Lett.}\ }\textbf {\bibinfo {volume} {119}},\
  \bibinfo {pages} {161101} (\bibinfo {year} {2017}{\natexlab{a}})},\ \Eprint
  {https://arxiv.org/abs/1710.05832} {arXiv:1710.05832 [gr-qc]} \BibitemShut
  {NoStop}%
\bibitem [{\citenamefont {Abbott}\ \emph
  {et~al.}(2016{\natexlab{b}})\citenamefont {Abbott} \emph
  {et~al.}}]{LIGOScientific:2016dsl}%
  \BibitemOpen
  \bibfield  {author} {\bibinfo {author} {\bibfnamefont {B.~P.}\ \bibnamefont
  {Abbott}} \emph {et~al.} (\bibinfo {collaboration} {LIGO Scientific,
  Virgo}),\ }\bibfield  {title} {\bibinfo {title} {{Binary Black Hole Mergers
  in the first Advanced LIGO Observing Run}},\ }\href
  {https://doi.org/10.1103/PhysRevX.6.041015} {\bibfield  {journal} {\bibinfo
  {journal} {Phys. Rev. X}\ }\textbf {\bibinfo {volume} {6}},\ \bibinfo {pages}
  {041015} (\bibinfo {year} {2016}{\natexlab{b}})},\ \bibinfo {note} {[Erratum:
  Phys.Rev.X 8, 039903 (2018)]},\ \Eprint {https://arxiv.org/abs/1606.04856}
  {arXiv:1606.04856 [gr-qc]} \BibitemShut {NoStop}%
\bibitem [{\citenamefont {Abbott}\ \emph
  {et~al.}(2019{\natexlab{a}})\citenamefont {Abbott} \emph
  {et~al.}}]{LIGOScientific:2018mvr}%
  \BibitemOpen
  \bibfield  {author} {\bibinfo {author} {\bibfnamefont {B.~P.}\ \bibnamefont
  {Abbott}} \emph {et~al.} (\bibinfo {collaboration} {LIGO Scientific,
  Virgo}),\ }\bibfield  {title} {\bibinfo {title} {{GWTC-1: A
  Gravitational-Wave Transient Catalog of Compact Binary Mergers Observed by
  LIGO and Virgo during the First and Second Observing Runs}},\ }\href
  {https://doi.org/10.1103/PhysRevX.9.031040} {\bibfield  {journal} {\bibinfo
  {journal} {Phys. Rev. X}\ }\textbf {\bibinfo {volume} {9}},\ \bibinfo {pages}
  {031040} (\bibinfo {year} {2019}{\natexlab{a}})},\ \Eprint
  {https://arxiv.org/abs/1811.12907} {arXiv:1811.12907 [astro-ph.HE]}
  \BibitemShut {NoStop}%
\bibitem [{\citenamefont {Abbott}\ \emph
  {et~al.}(2021{\natexlab{a}})\citenamefont {Abbott} \emph
  {et~al.}}]{LIGOScientific:2020ibl}%
  \BibitemOpen
  \bibfield  {author} {\bibinfo {author} {\bibfnamefont {R.}~\bibnamefont
  {Abbott}} \emph {et~al.} (\bibinfo {collaboration} {LIGO Scientific,
  Virgo}),\ }\bibfield  {title} {\bibinfo {title} {{GWTC-2: Compact Binary
  Coalescences Observed by LIGO and Virgo During the First Half of the Third
  Observing Run}},\ }\href {https://doi.org/10.1103/PhysRevX.11.021053}
  {\bibfield  {journal} {\bibinfo  {journal} {Phys. Rev. X}\ }\textbf {\bibinfo
  {volume} {11}},\ \bibinfo {pages} {021053} (\bibinfo {year}
  {2021}{\natexlab{a}})},\ \Eprint {https://arxiv.org/abs/2010.14527}
  {arXiv:2010.14527 [gr-qc]} \BibitemShut {NoStop}%
\bibitem [{\citenamefont {Abbott}\ \emph
  {et~al.}(2021{\natexlab{b}})\citenamefont {Abbott} \emph
  {et~al.}}]{LIGOScientific:2021usb}%
  \BibitemOpen
  \bibfield  {author} {\bibinfo {author} {\bibfnamefont {R.}~\bibnamefont
  {Abbott}} \emph {et~al.} (\bibinfo {collaboration} {LIGO Scientific,
  VIRGO}),\ }\bibfield  {title} {\bibinfo {title} {{GWTC-2.1: Deep Extended
  Catalog of Compact Binary Coalescences Observed by LIGO and Virgo During the
  First Half of the Third Observing Run}},\ }\href@noop {} {\  (\bibinfo {year}
  {2021}{\natexlab{b}})},\ \Eprint {https://arxiv.org/abs/2108.01045}
  {arXiv:2108.01045 [gr-qc]} \BibitemShut {NoStop}%
\bibitem [{\citenamefont {Abbott}\ \emph
  {et~al.}(2017{\natexlab{b}})\citenamefont {Abbott} \emph
  {et~al.}}]{TheLIGOScientific:2016dpb}%
  \BibitemOpen
  \bibfield  {author} {\bibinfo {author} {\bibfnamefont {B.~P.}\ \bibnamefont
  {Abbott}} \emph {et~al.} (\bibinfo {collaboration} {LIGO Scientific,
  Virgo}),\ }\bibfield  {title} {\bibinfo {title} {{Upper Limits on the
  Stochastic Gravitational-Wave Background from Advanced LIGO\textquoteright{}s
  First Observing Run}},\ }\href
  {https://doi.org/10.1103/PhysRevLett.118.121101} {\bibfield  {journal}
  {\bibinfo  {journal} {Phys. Rev. Lett.}\ }\textbf {\bibinfo {volume} {118}},\
  \bibinfo {pages} {121101} (\bibinfo {year} {2017}{\natexlab{b}})},\ \bibinfo
  {note} {[Erratum: Phys.Rev.Lett. 119, 029901 (2017)]},\ \Eprint
  {https://arxiv.org/abs/1612.02029} {arXiv:1612.02029 [gr-qc]} \BibitemShut
  {NoStop}%
\bibitem [{\citenamefont {Abbott}\ \emph {et~al.}(2018)\citenamefont {Abbott}
  \emph {et~al.}}]{Abbott:2018utx}%
  \BibitemOpen
  \bibfield  {author} {\bibinfo {author} {\bibfnamefont {B.~P.}\ \bibnamefont
  {Abbott}} \emph {et~al.} (\bibinfo {collaboration} {LIGO Scientific,
  Virgo}),\ }\bibfield  {title} {\bibinfo {title} {{Search for Tensor, Vector,
  and Scalar Polarizations in the Stochastic Gravitational-Wave Background}},\
  }\href {https://doi.org/10.1103/PhysRevLett.120.201102} {\bibfield  {journal}
  {\bibinfo  {journal} {Phys. Rev. Lett.}\ }\textbf {\bibinfo {volume} {120}},\
  \bibinfo {pages} {201102} (\bibinfo {year} {2018})},\ \Eprint
  {https://arxiv.org/abs/1802.10194} {arXiv:1802.10194 [gr-qc]} \BibitemShut
  {NoStop}%
\bibitem [{\citenamefont {Abbott}\ \emph
  {et~al.}(2019{\natexlab{b}})\citenamefont {Abbott} \emph
  {et~al.}}]{LIGOScientific:2019vic}%
  \BibitemOpen
  \bibfield  {author} {\bibinfo {author} {\bibfnamefont {B.~P.}\ \bibnamefont
  {Abbott}} \emph {et~al.} (\bibinfo {collaboration} {LIGO Scientific,
  Virgo}),\ }\bibfield  {title} {\bibinfo {title} {{Search for the isotropic
  stochastic background using data from Advanced LIGO\textquoteright{}s second
  observing run}},\ }\href {https://doi.org/10.1103/PhysRevD.100.061101}
  {\bibfield  {journal} {\bibinfo  {journal} {Phys. Rev. D}\ }\textbf {\bibinfo
  {volume} {100}},\ \bibinfo {pages} {061101} (\bibinfo {year}
  {2019}{\natexlab{b}})},\ \Eprint {https://arxiv.org/abs/1903.02886}
  {arXiv:1903.02886 [gr-qc]} \BibitemShut {NoStop}%
\bibitem [{\citenamefont {Abbott}\ \emph
  {et~al.}(2021{\natexlab{c}})\citenamefont {Abbott} \emph
  {et~al.}}]{Abbott:2021xxi}%
  \BibitemOpen
  \bibfield  {author} {\bibinfo {author} {\bibfnamefont {R.}~\bibnamefont
  {Abbott}} \emph {et~al.} (\bibinfo {collaboration} {LIGO Scientific, Virgo,
  KAGRA}),\ }\bibfield  {title} {\bibinfo {title} {{Upper Limits on the
  Isotropic Gravitational-Wave Background from Advanced LIGO's and Advanced
  Virgo's Third Observing Run}},\ }\href@noop {} {\  (\bibinfo {year}
  {2021}{\natexlab{c}})},\ \Eprint {https://arxiv.org/abs/2101.12130}
  {arXiv:2101.12130 [gr-qc]} \BibitemShut {NoStop}%
\bibitem [{\citenamefont {Abbott}\ \emph
  {et~al.}(2021{\natexlab{d}})\citenamefont {Abbott} \emph
  {et~al.}}]{Abbott:2021ksc}%
  \BibitemOpen
  \bibfield  {author} {\bibinfo {author} {\bibfnamefont {R.}~\bibnamefont
  {Abbott}} \emph {et~al.} (\bibinfo {collaboration} {LIGO Scientific, Virgo,
  KAGRA}),\ }\bibfield  {title} {\bibinfo {title} {{Constraints on cosmic
  strings using data from the third Advanced LIGO-Virgo observing run}},\
  }\href@noop {} {\  (\bibinfo {year} {2021}{\natexlab{d}})},\ \Eprint
  {https://arxiv.org/abs/2101.12248} {arXiv:2101.12248 [gr-qc]} \BibitemShut
  {NoStop}%
\bibitem [{\citenamefont {Abbott}\ \emph
  {et~al.}(2021{\natexlab{e}})\citenamefont {Abbott} \emph
  {et~al.}}]{Abbott:2021jel}%
  \BibitemOpen
  \bibfield  {author} {\bibinfo {author} {\bibfnamefont {R.}~\bibnamefont
  {Abbott}} \emph {et~al.} (\bibinfo {collaboration} {LIGO Scientific, Virgo,
  KAGRA}),\ }\bibfield  {title} {\bibinfo {title} {{Search for anisotropic
  gravitational-wave backgrounds using data from Advanced LIGO's and Advanced
  Virgo's first three observing runs}},\ }\href@noop {} {\  (\bibinfo {year}
  {2021}{\natexlab{e}})},\ \Eprint {https://arxiv.org/abs/2103.08520}
  {arXiv:2103.08520 [gr-qc]} \BibitemShut {NoStop}%
\bibitem [{\citenamefont {Arzoumanian}\ \emph {et~al.}(2020)\citenamefont
  {Arzoumanian} \emph {et~al.}}]{NANOGrav:2020bcs}%
  \BibitemOpen
  \bibfield  {author} {\bibinfo {author} {\bibfnamefont {Z.}~\bibnamefont
  {Arzoumanian}} \emph {et~al.} (\bibinfo {collaboration} {NANOGrav}),\
  }\bibfield  {title} {\bibinfo {title} {{The NANOGrav 12.5 yr Data Set: Search
  for an Isotropic Stochastic Gravitational-wave Background}},\ }\href
  {https://doi.org/10.3847/2041-8213/abd401} {\bibfield  {journal} {\bibinfo
  {journal} {Astrophys. J. Lett.}\ }\textbf {\bibinfo {volume} {905}},\
  \bibinfo {pages} {L34} (\bibinfo {year} {2020})},\ \Eprint
  {https://arxiv.org/abs/2009.04496} {arXiv:2009.04496 [astro-ph.HE]}
  \BibitemShut {NoStop}%
\bibitem [{\citenamefont {{Amaro-Seoane}}\ \emph {et~al.}(2017)\citenamefont
  {{Amaro-Seoane}}, \citenamefont {{Audley}}, \citenamefont {{Babak}},\ and\
  \citenamefont {{et al}}}]{2017arXiv170200786A}%
  \BibitemOpen
  \bibfield  {author} {\bibinfo {author} {\bibfnamefont {P.}~\bibnamefont
  {{Amaro-Seoane}}}, \bibinfo {author} {\bibfnamefont {H.}~\bibnamefont
  {{Audley}}}, \bibinfo {author} {\bibfnamefont {S.}~\bibnamefont {{Babak}}},\
  and\ \bibinfo {author} {\bibnamefont {{et al}}} (\bibinfo {collaboration}
  {{LISA Team}}),\ }\bibfield  {title} {\bibinfo {title} {{Laser Interferometer
  Space Antenna}},\ }\href@noop {} {\bibfield  {journal} {\bibinfo  {journal}
  {arXiv e-prints}\ ,\ \bibinfo {eid} {arXiv:1702.00786}} (\bibinfo {year}
  {2017})}\BibitemShut {NoStop}%
\bibitem [{\citenamefont {{LISA~Study~Team}}(2000)}]{LISA2000}%
  \BibitemOpen
  \bibfield  {author} {\bibinfo {author} {\bibnamefont {{LISA~Study~Team}}},\
  }\href@noop {} {\emph {\bibinfo {title} {LISA (Laser Interferometer Space
  Antenna): A Cornerstone Mission for the Observation of Gravitational
  Waves}}},\ \bibinfo {type} {Tech. Rep.}\ \bibinfo {number} {11}\ (\bibinfo
  {institution} {ESA-SCI},\ \bibinfo {year} {2000})\ \bibinfo {note} {system
  and Technology Study Report}\BibitemShut {NoStop}%
\bibitem [{\citenamefont {Hu}\ and\ \citenamefont {Wu}(2017)}]{Hu:2017mde}%
  \BibitemOpen
  \bibfield  {author} {\bibinfo {author} {\bibfnamefont {W.-R.}\ \bibnamefont
  {Hu}}\ and\ \bibinfo {author} {\bibfnamefont {Y.-L.}\ \bibnamefont {Wu}},\
  }\bibfield  {title} {\bibinfo {title} {{The Taiji Program in Space for
  gravitational wave physics and the nature of gravity}},\ }\href
  {https://doi.org/10.1093/nsr/nwx116} {\bibfield  {journal} {\bibinfo
  {journal} {Natl. Sci. Rev.}\ }\textbf {\bibinfo {volume} {4}},\ \bibinfo
  {pages} {685} (\bibinfo {year} {2017})}\BibitemShut {NoStop}%
\bibitem [{\citenamefont {Ruan}\ \emph {et~al.}(2020)\citenamefont {Ruan},
  \citenamefont {Liu}, \citenamefont {Guo}, \citenamefont {Wu},\ and\
  \citenamefont {Cai}}]{Ruan:2020smc}%
  \BibitemOpen
  \bibfield  {author} {\bibinfo {author} {\bibfnamefont {W.-H.}\ \bibnamefont
  {Ruan}}, \bibinfo {author} {\bibfnamefont {C.}~\bibnamefont {Liu}}, \bibinfo
  {author} {\bibfnamefont {Z.-K.}\ \bibnamefont {Guo}}, \bibinfo {author}
  {\bibfnamefont {Y.-L.}\ \bibnamefont {Wu}},\ and\ \bibinfo {author}
  {\bibfnamefont {R.-G.}\ \bibnamefont {Cai}},\ }\bibfield  {title} {\bibinfo
  {title} {{The LISA-Taiji network}},\ }\href
  {https://doi.org/10.1038/s41550-019-1008-4} {\bibfield  {journal} {\bibinfo
  {journal} {Nature Astron.}\ }\textbf {\bibinfo {volume} {4}},\ \bibinfo
  {pages} {108} (\bibinfo {year} {2020})},\ \Eprint
  {https://arxiv.org/abs/2002.03603} {arXiv:2002.03603 [gr-qc]} \BibitemShut
  {NoStop}%
\bibitem [{\citenamefont {{Wang}}\ \emph {et~al.}(2020)\citenamefont {{Wang}},
  \citenamefont {{Ni}}, \citenamefont {{Han}}, \citenamefont {{Yang}},\ and\
  \citenamefont {{Zhong}}}]{Wang:2020a}%
  \BibitemOpen
  \bibfield  {author} {\bibinfo {author} {\bibfnamefont {G.}~\bibnamefont
  {{Wang}}}, \bibinfo {author} {\bibfnamefont {W.-T.}\ \bibnamefont {{Ni}}},
  \bibinfo {author} {\bibfnamefont {W.-B.}\ \bibnamefont {{Han}}}, \bibinfo
  {author} {\bibfnamefont {S.-C.}\ \bibnamefont {{Yang}}},\ and\ \bibinfo
  {author} {\bibfnamefont {X.-Y.}\ \bibnamefont {{Zhong}}},\ }\bibfield
  {title} {\bibinfo {title} {{Numerical simulation of sky localization for
  LISA-TAIJI joint observation}},\ }\href
  {https://doi.org/10.1103/PhysRevD.102.024089} {\bibfield  {journal} {\bibinfo
   {journal} {\prd}\ }\textbf {\bibinfo {volume} {102}},\ \bibinfo {pages}
  {024089} (\bibinfo {year} {2020})},\ \Eprint
  {https://arxiv.org/abs/2002.12628} {arXiv:2002.12628} \BibitemShut {NoStop}%
\bibitem [{\citenamefont {Wang}\ and\ \citenamefont
  {Han}(2021)}]{Wang:2021polar}%
  \BibitemOpen
  \bibfield  {author} {\bibinfo {author} {\bibfnamefont {G.}~\bibnamefont
  {Wang}}\ and\ \bibinfo {author} {\bibfnamefont {W.-B.}\ \bibnamefont {Han}},\
  }\bibfield  {title} {\bibinfo {title} {{Observing gravitational wave
  polarizations with the LISA-TAIJI network}},\ }\href
  {https://doi.org/10.1103/PhysRevD.103.064021} {\bibfield  {journal} {\bibinfo
   {journal} {Phys. Rev. D}\ }\textbf {\bibinfo {volume} {103}},\ \bibinfo
  {pages} {064021} (\bibinfo {year} {2021})},\ \Eprint
  {https://arxiv.org/abs/2101.01991} {arXiv:2101.01991 [gr-qc]} \BibitemShut
  {NoStop}%
\bibitem [{\citenamefont {Shuman}\ and\ \citenamefont
  {Cornish}(2021)}]{Shuman:2021ruh}%
  \BibitemOpen
  \bibfield  {author} {\bibinfo {author} {\bibfnamefont {K.~J.}\ \bibnamefont
  {Shuman}}\ and\ \bibinfo {author} {\bibfnamefont {N.~J.}\ \bibnamefont
  {Cornish}},\ }\bibfield  {title} {\bibinfo {title} {{Massive Black Hole
  Binaries and Where to Find Them with Dual Detector Networks}},\ }\href@noop
  {} {\  (\bibinfo {year} {2021})},\ \Eprint {https://arxiv.org/abs/2105.02943}
  {arXiv:2105.02943 [gr-qc]} \BibitemShut {NoStop}%
\bibitem [{\citenamefont {Omiya}\ and\ \citenamefont
  {Seto}(2020)}]{Omiya:2020fvw}%
  \BibitemOpen
  \bibfield  {author} {\bibinfo {author} {\bibfnamefont {H.}~\bibnamefont
  {Omiya}}\ and\ \bibinfo {author} {\bibfnamefont {N.}~\bibnamefont {Seto}},\
  }\bibfield  {title} {\bibinfo {title} {{Searching for anomalous polarization
  modes of the stochastic gravitational wave background with LISA and Taiji}},\
  }\href {https://doi.org/10.1103/PhysRevD.102.084053} {\bibfield  {journal}
  {\bibinfo  {journal} {Phys. Rev. D}\ }\textbf {\bibinfo {volume} {102}},\
  \bibinfo {pages} {084053} (\bibinfo {year} {2020})},\ \Eprint
  {https://arxiv.org/abs/2010.00771} {arXiv:2010.00771 [gr-qc]} \BibitemShut
  {NoStop}%
\bibitem [{\citenamefont {Seto}(2020)}]{Seto:2020mfd}%
  \BibitemOpen
  \bibfield  {author} {\bibinfo {author} {\bibfnamefont {N.}~\bibnamefont
  {Seto}},\ }\bibfield  {title} {\bibinfo {title} {{Gravitational Wave
  Background Search by Correlating Multiple Triangular Detectors in the mHz
  Band}},\ }\href {https://doi.org/10.1103/PhysRevD.102.123547} {\bibfield
  {journal} {\bibinfo  {journal} {Phys. Rev. D}\ }\textbf {\bibinfo {volume}
  {102}},\ \bibinfo {pages} {123547} (\bibinfo {year} {2020})},\ \Eprint
  {https://arxiv.org/abs/2010.06877} {arXiv:2010.06877 [gr-qc]} \BibitemShut
  {NoStop}%
\bibitem [{\citenamefont {Orlando}\ \emph {et~al.}(2020)\citenamefont
  {Orlando}, \citenamefont {Pieroni},\ and\ \citenamefont
  {Ricciardone}}]{Orlando:2020oko}%
  \BibitemOpen
  \bibfield  {author} {\bibinfo {author} {\bibfnamefont {G.}~\bibnamefont
  {Orlando}}, \bibinfo {author} {\bibfnamefont {M.}~\bibnamefont {Pieroni}},\
  and\ \bibinfo {author} {\bibfnamefont {A.}~\bibnamefont {Ricciardone}},\
  }\bibfield  {title} {\bibinfo {title} {{Measuring Parity Violation in the
  Stochastic Gravitational Wave Background with the LISA-Taiji network}},\
  }\href@noop {} {\  (\bibinfo {year} {2020})},\ \Eprint
  {https://arxiv.org/abs/2011.07059} {arXiv:2011.07059 [astro-ph.CO]}
  \BibitemShut {NoStop}%
\bibitem [{\citenamefont {Pol}\ \emph {et~al.}(2021)\citenamefont {Pol},
  \citenamefont {Mandal}, \citenamefont {Brandenburg},\ and\ \citenamefont
  {Kahniashvili}}]{Pol:2021uol}%
  \BibitemOpen
  \bibfield  {author} {\bibinfo {author} {\bibfnamefont {A.~R.}\ \bibnamefont
  {Pol}}, \bibinfo {author} {\bibfnamefont {S.}~\bibnamefont {Mandal}},
  \bibinfo {author} {\bibfnamefont {A.}~\bibnamefont {Brandenburg}},\ and\
  \bibinfo {author} {\bibfnamefont {T.}~\bibnamefont {Kahniashvili}},\
  }\bibfield  {title} {\bibinfo {title} {{Polarization of gravitational waves
  from helical MHD turbulent sources}},\ }\href@noop {} {\  (\bibinfo {year}
  {2021})},\ \Eprint {https://arxiv.org/abs/2107.05356} {arXiv:2107.05356
  [gr-qc]} \BibitemShut {NoStop}%
\bibitem [{\citenamefont {Wang}\ \emph
  {et~al.}(2021{\natexlab{b}})\citenamefont {Wang}, \citenamefont {Jin},
  \citenamefont {Zhang},\ and\ \citenamefont {Zhang}}]{Wang:2021srv}%
  \BibitemOpen
  \bibfield  {author} {\bibinfo {author} {\bibfnamefont {L.-F.}\ \bibnamefont
  {Wang}}, \bibinfo {author} {\bibfnamefont {S.-J.}\ \bibnamefont {Jin}},
  \bibinfo {author} {\bibfnamefont {J.-F.}\ \bibnamefont {Zhang}},\ and\
  \bibinfo {author} {\bibfnamefont {X.}~\bibnamefont {Zhang}},\ }\bibfield
  {title} {\bibinfo {title} {{Forecast for cosmological parameter estimation
  with gravitational-wave standard sirens from the LISA-Taiji network}},\
  }\href@noop {} {\  (\bibinfo {year} {2021}{\natexlab{b}})},\ \Eprint
  {https://arxiv.org/abs/2101.11882} {arXiv:2101.11882 [gr-qc]} \BibitemShut
  {NoStop}%
\bibitem [{\citenamefont {Wang}\ \emph
  {et~al.}(2020{\natexlab{a}})\citenamefont {Wang}, \citenamefont {Ruan},
  \citenamefont {Yang}, \citenamefont {Guo}, \citenamefont {Cai},\ and\
  \citenamefont {Hu}}]{Wang:2020dkc}%
  \BibitemOpen
  \bibfield  {author} {\bibinfo {author} {\bibfnamefont {R.}~\bibnamefont
  {Wang}}, \bibinfo {author} {\bibfnamefont {W.-H.}\ \bibnamefont {Ruan}},
  \bibinfo {author} {\bibfnamefont {Q.}~\bibnamefont {Yang}}, \bibinfo {author}
  {\bibfnamefont {Z.-K.}\ \bibnamefont {Guo}}, \bibinfo {author} {\bibfnamefont
  {R.-G.}\ \bibnamefont {Cai}},\ and\ \bibinfo {author} {\bibfnamefont
  {B.}~\bibnamefont {Hu}},\ }\bibfield  {title} {\bibinfo {title} {{Hubble
  parameter estimation via dark sirens with the LISA-Taiji network}},\
  }\href@noop {} {\  (\bibinfo {year} {2020}{\natexlab{a}})},\ \Eprint
  {https://arxiv.org/abs/2010.14732} {arXiv:2010.14732 [astro-ph.CO]}
  \BibitemShut {NoStop}%
\bibitem [{\citenamefont {Korol}\ \emph {et~al.}(2017)\citenamefont {Korol},
  \citenamefont {Rossi}, \citenamefont {Groot}, \citenamefont {Nelemans},
  \citenamefont {Toonen},\ and\ \citenamefont {Brown}}]{Korol:2017qcx}%
  \BibitemOpen
  \bibfield  {author} {\bibinfo {author} {\bibfnamefont {V.}~\bibnamefont
  {Korol}}, \bibinfo {author} {\bibfnamefont {E.~M.}\ \bibnamefont {Rossi}},
  \bibinfo {author} {\bibfnamefont {P.~J.}\ \bibnamefont {Groot}}, \bibinfo
  {author} {\bibfnamefont {G.}~\bibnamefont {Nelemans}}, \bibinfo {author}
  {\bibfnamefont {S.}~\bibnamefont {Toonen}},\ and\ \bibinfo {author}
  {\bibfnamefont {A.~G.~A.}\ \bibnamefont {Brown}},\ }\bibfield  {title}
  {\bibinfo {title} {{Prospects for detection of detached double white dwarf
  binaries with Gaia, LSST and LISA}},\ }\href
  {https://doi.org/10.1093/mnras/stx1285} {\bibfield  {journal} {\bibinfo
  {journal} {Mon. Not. Roy. Astron. Soc.}\ }\textbf {\bibinfo {volume} {470}},\
  \bibinfo {pages} {1894} (\bibinfo {year} {2017})},\ \Eprint
  {https://arxiv.org/abs/1703.02555} {arXiv:1703.02555 [astro-ph.HE]}
  \BibitemShut {NoStop}%
\bibitem [{\citenamefont {Cornish}\ and\ \citenamefont
  {Robson}(2017)}]{Cornish:2017vip}%
  \BibitemOpen
  \bibfield  {author} {\bibinfo {author} {\bibfnamefont {N.}~\bibnamefont
  {Cornish}}\ and\ \bibinfo {author} {\bibfnamefont {T.}~\bibnamefont
  {Robson}},\ }\bibfield  {title} {\bibinfo {title} {{Galactic binary science
  with the new LISA design}},\ }\href
  {https://doi.org/10.1088/1742-6596/840/1/012024} {\bibfield  {journal}
  {\bibinfo  {journal} {J. Phys. Conf. Ser.}\ }\textbf {\bibinfo {volume}
  {840}},\ \bibinfo {pages} {012024} (\bibinfo {year} {2017})},\ \Eprint
  {https://arxiv.org/abs/1703.09858} {arXiv:1703.09858 [astro-ph.IM]}
  \BibitemShut {NoStop}%
\bibitem [{\citenamefont {Romano}\ and\ \citenamefont
  {Cornish}(2017)}]{Romano:2016dpx}%
  \BibitemOpen
  \bibfield  {author} {\bibinfo {author} {\bibfnamefont {J.~D.}\ \bibnamefont
  {Romano}}\ and\ \bibinfo {author} {\bibfnamefont {N.~J.}\ \bibnamefont
  {Cornish}},\ }\bibfield  {title} {\bibinfo {title} {{Detection methods for
  stochastic gravitational-wave backgrounds: a unified treatment}},\ }\href
  {https://doi.org/10.1007/s41114-017-0004-1} {\bibfield  {journal} {\bibinfo
  {journal} {Living Rev. Rel.}\ }\textbf {\bibinfo {volume} {20}},\ \bibinfo
  {pages} {2} (\bibinfo {year} {2017})},\ \Eprint
  {https://arxiv.org/abs/1608.06889} {arXiv:1608.06889 [gr-qc]} \BibitemShut
  {NoStop}%
\bibitem [{\citenamefont {Caprini}\ \emph {et~al.}(2016)\citenamefont {Caprini}
  \emph {et~al.}}]{Caprini:2015zlo}%
  \BibitemOpen
  \bibfield  {author} {\bibinfo {author} {\bibfnamefont {C.}~\bibnamefont
  {Caprini}} \emph {et~al.},\ }\bibfield  {title} {\bibinfo {title} {{Science
  with the space-based interferometer eLISA. II: Gravitational waves from
  cosmological phase transitions}},\ }\href
  {https://doi.org/10.1088/1475-7516/2016/04/001} {\bibfield  {journal}
  {\bibinfo  {journal} {JCAP}\ }\textbf {\bibinfo {volume} {04}},\ \bibinfo
  {pages} {001}},\ \Eprint {https://arxiv.org/abs/1512.06239} {arXiv:1512.06239
  [astro-ph.CO]} \BibitemShut {NoStop}%
\bibitem [{\citenamefont {Bartolo}\ \emph {et~al.}(2016)\citenamefont {Bartolo}
  \emph {et~al.}}]{Bartolo:2016ami}%
  \BibitemOpen
  \bibfield  {author} {\bibinfo {author} {\bibfnamefont {N.}~\bibnamefont
  {Bartolo}} \emph {et~al.},\ }\bibfield  {title} {\bibinfo {title} {{Science
  with the space-based interferometer LISA. IV: Probing inflation with
  gravitational waves}},\ }\href
  {https://doi.org/10.1088/1475-7516/2016/12/026} {\bibfield  {journal}
  {\bibinfo  {journal} {JCAP}\ }\textbf {\bibinfo {volume} {12}},\ \bibinfo
  {pages} {026}},\ \Eprint {https://arxiv.org/abs/1610.06481} {arXiv:1610.06481
  [astro-ph.CO]} \BibitemShut {NoStop}%
\bibitem [{\citenamefont {Caprini}\ \emph {et~al.}(2020)\citenamefont {Caprini}
  \emph {et~al.}}]{Caprini:2019egz}%
  \BibitemOpen
  \bibfield  {author} {\bibinfo {author} {\bibfnamefont {C.}~\bibnamefont
  {Caprini}} \emph {et~al.},\ }\bibfield  {title} {\bibinfo {title} {{Detecting
  gravitational waves from cosmological phase transitions with LISA: an
  update}},\ }\href {https://doi.org/10.1088/1475-7516/2020/03/024} {\bibfield
  {journal} {\bibinfo  {journal} {JCAP}\ }\textbf {\bibinfo {volume} {03}},\
  \bibinfo {pages} {024}},\ \Eprint {https://arxiv.org/abs/1910.13125}
  {arXiv:1910.13125 [astro-ph.CO]} \BibitemShut {NoStop}%
\bibitem [{\citenamefont {Caprini}\ \emph {et~al.}(2019)\citenamefont
  {Caprini}, \citenamefont {Figueroa}, \citenamefont {Flauger}, \citenamefont
  {Nardini}, \citenamefont {Peloso}, \citenamefont {Pieroni}, \citenamefont
  {Ricciardone},\ and\ \citenamefont {Tasinato}}]{Caprini:2019pxz}%
  \BibitemOpen
  \bibfield  {author} {\bibinfo {author} {\bibfnamefont {C.}~\bibnamefont
  {Caprini}}, \bibinfo {author} {\bibfnamefont {D.~G.}\ \bibnamefont
  {Figueroa}}, \bibinfo {author} {\bibfnamefont {R.}~\bibnamefont {Flauger}},
  \bibinfo {author} {\bibfnamefont {G.}~\bibnamefont {Nardini}}, \bibinfo
  {author} {\bibfnamefont {M.}~\bibnamefont {Peloso}}, \bibinfo {author}
  {\bibfnamefont {M.}~\bibnamefont {Pieroni}}, \bibinfo {author} {\bibfnamefont
  {A.}~\bibnamefont {Ricciardone}},\ and\ \bibinfo {author} {\bibfnamefont
  {G.}~\bibnamefont {Tasinato}},\ }\bibfield  {title} {\bibinfo {title}
  {{Reconstructing the spectral shape of a stochastic gravitational wave
  background with LISA}},\ }\href
  {https://doi.org/10.1088/1475-7516/2019/11/017} {\bibfield  {journal}
  {\bibinfo  {journal} {JCAP}\ }\textbf {\bibinfo {volume} {11}},\ \bibinfo
  {pages} {017}},\ \Eprint {https://arxiv.org/abs/1906.09244} {arXiv:1906.09244
  [astro-ph.CO]} \BibitemShut {NoStop}%
\bibitem [{\citenamefont {Flauger}\ \emph {et~al.}(2021)\citenamefont
  {Flauger}, \citenamefont {Karnesis}, \citenamefont {Nardini}, \citenamefont
  {Pieroni}, \citenamefont {Ricciardone},\ and\ \citenamefont
  {Torrado}}]{Flauger:2020qyi}%
  \BibitemOpen
  \bibfield  {author} {\bibinfo {author} {\bibfnamefont {R.}~\bibnamefont
  {Flauger}}, \bibinfo {author} {\bibfnamefont {N.}~\bibnamefont {Karnesis}},
  \bibinfo {author} {\bibfnamefont {G.}~\bibnamefont {Nardini}}, \bibinfo
  {author} {\bibfnamefont {M.}~\bibnamefont {Pieroni}}, \bibinfo {author}
  {\bibfnamefont {A.}~\bibnamefont {Ricciardone}},\ and\ \bibinfo {author}
  {\bibfnamefont {J.}~\bibnamefont {Torrado}},\ }\bibfield  {title} {\bibinfo
  {title} {{Improved reconstruction of a stochastic gravitational wave
  background with LISA}},\ }\href
  {https://doi.org/10.1088/1475-7516/2021/01/059} {\bibfield  {journal}
  {\bibinfo  {journal} {JCAP}\ }\textbf {\bibinfo {volume} {01}},\ \bibinfo
  {pages} {059}},\ \Eprint {https://arxiv.org/abs/2009.11845} {arXiv:2009.11845
  [astro-ph.CO]} \BibitemShut {NoStop}%
\bibitem [{\citenamefont {Christensen}(2019)}]{Christensen:2018iqi}%
  \BibitemOpen
  \bibfield  {author} {\bibinfo {author} {\bibfnamefont {N.}~\bibnamefont
  {Christensen}},\ }\bibfield  {title} {\bibinfo {title} {{Stochastic
  Gravitational Wave Backgrounds}},\ }\href
  {https://doi.org/10.1088/1361-6633/aae6b5} {\bibfield  {journal} {\bibinfo
  {journal} {Rept. Prog. Phys.}\ }\textbf {\bibinfo {volume} {82}},\ \bibinfo
  {pages} {016903} (\bibinfo {year} {2019})},\ \Eprint
  {https://arxiv.org/abs/1811.08797} {arXiv:1811.08797 [gr-qc]} \BibitemShut
  {NoStop}%
\bibitem [{\citenamefont {{Estabrook}}\ and\ \citenamefont
  {{Wahlquist}}(1975)}]{1975GReGr...6..439E}%
  \BibitemOpen
  \bibfield  {author} {\bibinfo {author} {\bibfnamefont {F.~B.}\ \bibnamefont
  {{Estabrook}}}\ and\ \bibinfo {author} {\bibfnamefont {H.~D.}\ \bibnamefont
  {{Wahlquist}}},\ }\bibfield  {title} {\bibinfo {title} {{Response of Doppler
  spacecraft tracking to gravitational radiation.}},\ }\href
  {https://doi.org/10.1007/BF00762449} {\bibfield  {journal} {\bibinfo
  {journal} {General Relativity and Gravitation}\ }\textbf {\bibinfo {volume}
  {6}},\ \bibinfo {pages} {439} (\bibinfo {year} {1975})}\BibitemShut {NoStop}%
\bibitem [{\citenamefont {{Wahlquist}}(1987)}]{1987GReGr..19.1101W}%
  \BibitemOpen
  \bibfield  {author} {\bibinfo {author} {\bibfnamefont {H.}~\bibnamefont
  {{Wahlquist}}},\ }\bibfield  {title} {\bibinfo {title} {{The Doppler response
  to gravitational waves from a binary star source.}},\ }\href
  {https://doi.org/10.1007/BF00759146} {\bibfield  {journal} {\bibinfo
  {journal} {General Relativity and Gravitation}\ }\textbf {\bibinfo {volume}
  {19}},\ \bibinfo {pages} {1101} (\bibinfo {year} {1987})}\BibitemShut
  {NoStop}%
\bibitem [{\citenamefont {Vallisneri}\ \emph {et~al.}(2008)\citenamefont
  {Vallisneri}, \citenamefont {Crowder},\ and\ \citenamefont
  {Tinto}}]{Vallisneri:2007xa}%
  \BibitemOpen
  \bibfield  {author} {\bibinfo {author} {\bibfnamefont {M.}~\bibnamefont
  {Vallisneri}}, \bibinfo {author} {\bibfnamefont {J.}~\bibnamefont
  {Crowder}},\ and\ \bibinfo {author} {\bibfnamefont {M.}~\bibnamefont
  {Tinto}},\ }\bibfield  {title} {\bibinfo {title} {{Sensitivity and
  parameter-estimation precision for alternate LISA configurations}},\ }\href
  {https://doi.org/10.1088/0264-9381/25/6/065005} {\bibfield  {journal}
  {\bibinfo  {journal} {Class. Quant. Grav.}\ }\textbf {\bibinfo {volume}
  {25}},\ \bibinfo {pages} {065005} (\bibinfo {year} {2008})},\ \Eprint
  {https://arxiv.org/abs/0710.4369} {arXiv:0710.4369 [gr-qc]} \BibitemShut
  {NoStop}%
\bibitem [{\citenamefont {Vallisneri}\ and\ \citenamefont
  {Galley}(2012)}]{Vallisneri:2012np}%
  \BibitemOpen
  \bibfield  {author} {\bibinfo {author} {\bibfnamefont {M.}~\bibnamefont
  {Vallisneri}}\ and\ \bibinfo {author} {\bibfnamefont {C.~R.}\ \bibnamefont
  {Galley}},\ }\bibfield  {title} {\bibinfo {title} {{Non-sky-averaged
  sensitivity curves for space-based gravitational-wave observatories}},\
  }\href {https://doi.org/10.1088/0264-9381/29/12/124015} {\bibfield  {journal}
  {\bibinfo  {journal} {Class. Quant. Grav.}\ }\textbf {\bibinfo {volume}
  {29}},\ \bibinfo {pages} {124015} (\bibinfo {year} {2012})},\ \Eprint
  {https://arxiv.org/abs/1201.3684} {arXiv:1201.3684 [gr-qc]} \BibitemShut
  {NoStop}%
\bibitem [{\citenamefont {Tinto}\ and\ \citenamefont
  {da~Silva~Alves}(2010)}]{Tinto:2010hz}%
  \BibitemOpen
  \bibfield  {author} {\bibinfo {author} {\bibfnamefont {M.}~\bibnamefont
  {Tinto}}\ and\ \bibinfo {author} {\bibfnamefont {M.~E.}\ \bibnamefont
  {da~Silva~Alves}},\ }\bibfield  {title} {\bibinfo {title} {{LISA
  Sensitivities to Gravitational Waves from Relativistic Metric Theories of
  Gravity}},\ }\href {https://doi.org/10.1103/PhysRevD.82.122003} {\bibfield
  {journal} {\bibinfo  {journal} {Phys. Rev. D}\ }\textbf {\bibinfo {volume}
  {82}},\ \bibinfo {pages} {122003} (\bibinfo {year} {2010})},\ \Eprint
  {https://arxiv.org/abs/1010.1302} {arXiv:1010.1302 [gr-qc]} \BibitemShut
  {NoStop}%
\bibitem [{\citenamefont {Otto}\ \emph {et~al.}(2012)\citenamefont {Otto},
  \citenamefont {Heinzel},\ and\ \citenamefont {Danzmann}}]{Otto:2012dk}%
  \BibitemOpen
  \bibfield  {author} {\bibinfo {author} {\bibfnamefont {M.}~\bibnamefont
  {Otto}}, \bibinfo {author} {\bibfnamefont {G.}~\bibnamefont {Heinzel}},\ and\
  \bibinfo {author} {\bibfnamefont {K.}~\bibnamefont {Danzmann}},\ }\bibfield
  {title} {\bibinfo {title} {{TDI and clock noise removal for the split
  interferometry configuration of LISA}},\ }\href
  {https://doi.org/10.1088/0264-9381/29/20/205003} {\bibfield  {journal}
  {\bibinfo  {journal} {Class. Quant. Grav.}\ }\textbf {\bibinfo {volume}
  {29}},\ \bibinfo {pages} {205003} (\bibinfo {year} {2012})}\BibitemShut
  {NoStop}%
\bibitem [{\citenamefont {Otto}(2015)}]{Otto:2015}%
  \BibitemOpen
  \bibfield  {author} {\bibinfo {author} {\bibfnamefont {M.}~\bibnamefont
  {Otto}},\ }\emph {\bibinfo {title} {{Time-Delay Interferometry Simulations
  for the Laser Interferometer Space Antenna}}},\ \href
  {https://doi.org/10.15488/8545} {Ph.D. thesis},\ \bibinfo  {school} {Leibniz
  U., Hannover} (\bibinfo {year} {2015})\BibitemShut {NoStop}%
\bibitem [{\citenamefont {Tinto}\ and\ \citenamefont
  {Hartwig}(2018)}]{Tinto:2018kij}%
  \BibitemOpen
  \bibfield  {author} {\bibinfo {author} {\bibfnamefont {M.}~\bibnamefont
  {Tinto}}\ and\ \bibinfo {author} {\bibfnamefont {O.}~\bibnamefont
  {Hartwig}},\ }\bibfield  {title} {\bibinfo {title} {{Time-Delay
  Interferometry and Clock-Noise Calibration}},\ }\href
  {https://doi.org/10.1103/PhysRevD.98.042003} {\bibfield  {journal} {\bibinfo
  {journal} {Phys. Rev. D}\ }\textbf {\bibinfo {volume} {98}},\ \bibinfo
  {pages} {042003} (\bibinfo {year} {2018})},\ \Eprint
  {https://arxiv.org/abs/1807.02594} {arXiv:1807.02594 [gr-qc]} \BibitemShut
  {NoStop}%
\bibitem [{\citenamefont {Wang}\ \emph
  {et~al.}(2020{\natexlab{b}})\citenamefont {Wang}, \citenamefont {Ni},
  \citenamefont {Han},\ and\ \citenamefont {Qiao}}]{Wang:2ndTDI}%
  \BibitemOpen
  \bibfield  {author} {\bibinfo {author} {\bibfnamefont {G.}~\bibnamefont
  {Wang}}, \bibinfo {author} {\bibfnamefont {W.-T.}\ \bibnamefont {Ni}},
  \bibinfo {author} {\bibfnamefont {W.-B.}\ \bibnamefont {Han}},\ and\ \bibinfo
  {author} {\bibfnamefont {C.-F.}\ \bibnamefont {Qiao}},\ }\bibfield  {title}
  {\bibinfo {title} {{Algorithm for TDI numerical simulation and sensitivity
  investigation}},\ }\href@noop {} {\  (\bibinfo {year}
  {2020}{\natexlab{b}})},\ \Eprint {https://arxiv.org/abs/2010.15544}
  {arXiv:2010.15544 [gr-qc]} \BibitemShut {NoStop}%
\bibitem [{\citenamefont {Prince}\ \emph {et~al.}(2002)\citenamefont {Prince},
  \citenamefont {Tinto}, \citenamefont {Larson},\ and\ \citenamefont
  {Armstrong}}]{Prince:2002hp}%
  \BibitemOpen
  \bibfield  {author} {\bibinfo {author} {\bibfnamefont {T.~A.}\ \bibnamefont
  {Prince}}, \bibinfo {author} {\bibfnamefont {M.}~\bibnamefont {Tinto}},
  \bibinfo {author} {\bibfnamefont {S.~L.}\ \bibnamefont {Larson}},\ and\
  \bibinfo {author} {\bibfnamefont {J.~W.}\ \bibnamefont {Armstrong}},\
  }\bibfield  {title} {\bibinfo {title} {{The LISA optimal sensitivity}},\
  }\href {https://doi.org/10.1103/PhysRevD.66.122002} {\bibfield  {journal}
  {\bibinfo  {journal} {Phys. Rev. D}\ }\textbf {\bibinfo {volume} {66}},\
  \bibinfo {pages} {122002} (\bibinfo {year} {2002})},\ \Eprint
  {https://arxiv.org/abs/gr-qc/0209039} {arXiv:gr-qc/0209039 [gr-qc]}
  \BibitemShut {NoStop}%
\bibitem [{\citenamefont {Luo}\ \emph {et~al.}(2020)\citenamefont {Luo},
  \citenamefont {Guo}, \citenamefont {Jin}, \citenamefont {Wu},\ and\
  \citenamefont {Hu}}]{Luo:2020}%
  \BibitemOpen
  \bibfield  {author} {\bibinfo {author} {\bibfnamefont {Z.}~\bibnamefont
  {Luo}}, \bibinfo {author} {\bibfnamefont {Z.}~\bibnamefont {Guo}}, \bibinfo
  {author} {\bibfnamefont {G.}~\bibnamefont {Jin}}, \bibinfo {author}
  {\bibfnamefont {Y.}~\bibnamefont {Wu}},\ and\ \bibinfo {author}
  {\bibfnamefont {W.}~\bibnamefont {Hu}},\ }\bibfield  {title} {\bibinfo
  {title} {{A brief analysis to Taiji: Science and technology}},\ }\href
  {https://doi.org/doi.org/10.1016/j.rinp.2019.102918} {\bibfield  {journal}
  {\bibinfo  {journal} {Results in Physics}\ }\textbf {\bibinfo {volume}
  {16}},\ \bibinfo {pages} {102918} (\bibinfo {year} {2020})}\BibitemShut
  {NoStop}%
\bibitem [{\citenamefont {Wang}\ \emph
  {et~al.}(2020{\natexlab{c}})\citenamefont {Wang}, \citenamefont {Ni},\ and\
  \citenamefont {Han}}]{Wang:1stTDI}%
  \BibitemOpen
  \bibfield  {author} {\bibinfo {author} {\bibfnamefont {G.}~\bibnamefont
  {Wang}}, \bibinfo {author} {\bibfnamefont {W.-T.}\ \bibnamefont {Ni}},\ and\
  \bibinfo {author} {\bibfnamefont {W.-B.}\ \bibnamefont {Han}},\ }\bibfield
  {title} {\bibinfo {title} {{Revisiting time delay interferometry for
  unequal-arm LISA and TAIJI}},\ }\href@noop {} {\  (\bibinfo {year}
  {2020}{\natexlab{c}})},\ \Eprint {https://arxiv.org/abs/2008.05812}
  {arXiv:2008.05812 [gr-qc]} \BibitemShut {NoStop}%
\bibitem [{\citenamefont {Allen}(1996)}]{Allen:1996vm}%
  \BibitemOpen
  \bibfield  {author} {\bibinfo {author} {\bibfnamefont {B.}~\bibnamefont
  {Allen}},\ }\bibfield  {title} {\bibinfo {title} {{The Stochastic gravity
  wave background: Sources and detection}},\ }in\ \href@noop {} {\emph
  {\bibinfo {booktitle} {{Les Houches School of Physics: Astrophysical Sources
  of Gravitational Radiation}}}}\ (\bibinfo {year} {1996})\ \Eprint
  {https://arxiv.org/abs/gr-qc/9604033} {arXiv:gr-qc/9604033} \BibitemShut
  {NoStop}%
\bibitem [{\citenamefont {Allen}\ and\ \citenamefont
  {Romano}(1999)}]{Allen:1997ad}%
  \BibitemOpen
  \bibfield  {author} {\bibinfo {author} {\bibfnamefont {B.}~\bibnamefont
  {Allen}}\ and\ \bibinfo {author} {\bibfnamefont {J.~D.}\ \bibnamefont
  {Romano}},\ }\bibfield  {title} {\bibinfo {title} {{Detecting a stochastic
  background of gravitational radiation: Signal processing strategies and
  sensitivities}},\ }\href {https://doi.org/10.1103/PhysRevD.59.102001}
  {\bibfield  {journal} {\bibinfo  {journal} {Phys. Rev. D}\ }\textbf {\bibinfo
  {volume} {59}},\ \bibinfo {pages} {102001} (\bibinfo {year} {1999})},\
  \Eprint {https://arxiv.org/abs/gr-qc/9710117} {arXiv:gr-qc/9710117}
  \BibitemShut {NoStop}%
\bibitem [{\citenamefont {Aghanim}\ \emph {et~al.}(2020)\citenamefont {Aghanim}
  \emph {et~al.}}]{Planck:2018vyg}%
  \BibitemOpen
  \bibfield  {author} {\bibinfo {author} {\bibfnamefont {N.}~\bibnamefont
  {Aghanim}} \emph {et~al.} (\bibinfo {collaboration} {Planck}),\ }\bibfield
  {title} {\bibinfo {title} {{Planck 2018 results. VI. Cosmological
  parameters}},\ }\href {https://doi.org/10.1051/0004-6361/201833910}
  {\bibfield  {journal} {\bibinfo  {journal} {Astron. Astrophys.}\ }\textbf
  {\bibinfo {volume} {641}},\ \bibinfo {pages} {A6} (\bibinfo {year} {2020})},\
  \bibinfo {note} {[Erratum: Astron.Astrophys. 652, C4 (2021)]},\ \Eprint
  {https://arxiv.org/abs/1807.06209} {arXiv:1807.06209 [astro-ph.CO]}
  \BibitemShut {NoStop}%
\bibitem [{\citenamefont {Kuroyanagi}\ \emph {et~al.}(2018)\citenamefont
  {Kuroyanagi}, \citenamefont {Chiba},\ and\ \citenamefont
  {Takahashi}}]{Kuroyanagi:2018csn}%
  \BibitemOpen
  \bibfield  {author} {\bibinfo {author} {\bibfnamefont {S.}~\bibnamefont
  {Kuroyanagi}}, \bibinfo {author} {\bibfnamefont {T.}~\bibnamefont {Chiba}},\
  and\ \bibinfo {author} {\bibfnamefont {T.}~\bibnamefont {Takahashi}},\
  }\bibfield  {title} {\bibinfo {title} {{Probing the Universe through the
  Stochastic Gravitational Wave Background}},\ }\href
  {https://doi.org/10.1088/1475-7516/2018/11/038} {\bibfield  {journal}
  {\bibinfo  {journal} {JCAP}\ }\textbf {\bibinfo {volume} {11}},\ \bibinfo
  {pages} {038}},\ \Eprint {https://arxiv.org/abs/1807.00786} {arXiv:1807.00786
  [astro-ph.CO]} \BibitemShut {NoStop}%
\bibitem [{\citenamefont {Martinovic}\ \emph {et~al.}(2021)\citenamefont
  {Martinovic}, \citenamefont {Meyers}, \citenamefont {Sakellariadou},\ and\
  \citenamefont {Christensen}}]{Martinovic:2020hru}%
  \BibitemOpen
  \bibfield  {author} {\bibinfo {author} {\bibfnamefont {K.}~\bibnamefont
  {Martinovic}}, \bibinfo {author} {\bibfnamefont {P.~M.}\ \bibnamefont
  {Meyers}}, \bibinfo {author} {\bibfnamefont {M.}~\bibnamefont
  {Sakellariadou}},\ and\ \bibinfo {author} {\bibfnamefont {N.}~\bibnamefont
  {Christensen}},\ }\bibfield  {title} {\bibinfo {title} {{Simultaneous
  estimation of astrophysical and cosmological stochastic gravitational-wave
  backgrounds with terrestrial detectors}},\ }\href
  {https://doi.org/10.1103/PhysRevD.103.043023} {\bibfield  {journal} {\bibinfo
   {journal} {Phys. Rev. D}\ }\textbf {\bibinfo {volume} {103}},\ \bibinfo
  {pages} {043023} (\bibinfo {year} {2021})},\ \Eprint
  {https://arxiv.org/abs/2011.05697} {arXiv:2011.05697 [gr-qc]} \BibitemShut
  {NoStop}%
\bibitem [{\citenamefont {Cornish}\ and\ \citenamefont
  {Larson}(2001)}]{Cornish:2001qi}%
  \BibitemOpen
  \bibfield  {author} {\bibinfo {author} {\bibfnamefont {N.~J.}\ \bibnamefont
  {Cornish}}\ and\ \bibinfo {author} {\bibfnamefont {S.~L.}\ \bibnamefont
  {Larson}},\ }\bibfield  {title} {\bibinfo {title} {{Space missions to detect
  the cosmic gravitational wave background}},\ }\href
  {https://doi.org/10.1088/0264-9381/18/17/308} {\bibfield  {journal} {\bibinfo
   {journal} {Class. Quant. Grav.}\ }\textbf {\bibinfo {volume} {18}},\
  \bibinfo {pages} {3473} (\bibinfo {year} {2001})},\ \Eprint
  {https://arxiv.org/abs/gr-qc/0103075} {arXiv:gr-qc/0103075} \BibitemShut
  {NoStop}%
\bibitem [{\citenamefont {Hindmarsh}\ \emph {et~al.}(2014)\citenamefont
  {Hindmarsh}, \citenamefont {Huber}, \citenamefont {Rummukainen},\ and\
  \citenamefont {Weir}}]{Hindmarsh:2013xza}%
  \BibitemOpen
  \bibfield  {author} {\bibinfo {author} {\bibfnamefont {M.}~\bibnamefont
  {Hindmarsh}}, \bibinfo {author} {\bibfnamefont {S.~J.}\ \bibnamefont
  {Huber}}, \bibinfo {author} {\bibfnamefont {K.}~\bibnamefont {Rummukainen}},\
  and\ \bibinfo {author} {\bibfnamefont {D.~J.}\ \bibnamefont {Weir}},\
  }\bibfield  {title} {\bibinfo {title} {{Gravitational waves from the sound of
  a first order phase transition}},\ }\href
  {https://doi.org/10.1103/PhysRevLett.112.041301} {\bibfield  {journal}
  {\bibinfo  {journal} {Phys. Rev. Lett.}\ }\textbf {\bibinfo {volume} {112}},\
  \bibinfo {pages} {041301} (\bibinfo {year} {2014})},\ \Eprint
  {https://arxiv.org/abs/1304.2433} {arXiv:1304.2433 [hep-ph]} \BibitemShut
  {NoStop}%
\bibitem [{\citenamefont {Thrane}\ and\ \citenamefont
  {Romano}(2013)}]{Thrane:2013oya}%
  \BibitemOpen
  \bibfield  {author} {\bibinfo {author} {\bibfnamefont {E.}~\bibnamefont
  {Thrane}}\ and\ \bibinfo {author} {\bibfnamefont {J.~D.}\ \bibnamefont
  {Romano}},\ }\bibfield  {title} {\bibinfo {title} {{Sensitivity curves for
  searches for gravitational-wave backgrounds}},\ }\href
  {https://doi.org/10.1103/PhysRevD.88.124032} {\bibfield  {journal} {\bibinfo
  {journal} {Phys. Rev. D}\ }\textbf {\bibinfo {volume} {88}},\ \bibinfo
  {pages} {124032} (\bibinfo {year} {2013})},\ \Eprint
  {https://arxiv.org/abs/1310.5300} {arXiv:1310.5300 [astro-ph.IM]}
  \BibitemShut {NoStop}%
\bibitem [{\citenamefont {Adams}\ and\ \citenamefont
  {Cornish}(2010)}]{Adams:2010vc}%
  \BibitemOpen
  \bibfield  {author} {\bibinfo {author} {\bibfnamefont {M.~R.}\ \bibnamefont
  {Adams}}\ and\ \bibinfo {author} {\bibfnamefont {N.~J.}\ \bibnamefont
  {Cornish}},\ }\bibfield  {title} {\bibinfo {title} {{Discriminating between a
  Stochastic Gravitational Wave Background and Instrument Noise}},\ }\href
  {https://doi.org/10.1103/PhysRevD.82.022002} {\bibfield  {journal} {\bibinfo
  {journal} {Phys. Rev. D}\ }\textbf {\bibinfo {volume} {82}},\ \bibinfo
  {pages} {022002} (\bibinfo {year} {2010})},\ \Eprint
  {https://arxiv.org/abs/1002.1291} {arXiv:1002.1291 [gr-qc]} \BibitemShut
  {NoStop}%
\bibitem [{\citenamefont {Adams}\ and\ \citenamefont
  {Cornish}(2014)}]{Adams:2013qma}%
  \BibitemOpen
  \bibfield  {author} {\bibinfo {author} {\bibfnamefont {M.~R.}\ \bibnamefont
  {Adams}}\ and\ \bibinfo {author} {\bibfnamefont {N.~J.}\ \bibnamefont
  {Cornish}},\ }\bibfield  {title} {\bibinfo {title} {{Detecting a Stochastic
  Gravitational Wave Background in the presence of a Galactic Foreground and
  Instrument Noise}},\ }\href {https://doi.org/10.1103/PhysRevD.89.022001}
  {\bibfield  {journal} {\bibinfo  {journal} {Phys. Rev. D}\ }\textbf {\bibinfo
  {volume} {89}},\ \bibinfo {pages} {022001} (\bibinfo {year} {2014})},\
  \Eprint {https://arxiv.org/abs/1307.4116} {arXiv:1307.4116 [gr-qc]}
  \BibitemShut {NoStop}%
\bibitem [{\citenamefont {Boileau}\ \emph {et~al.}(2021)\citenamefont
  {Boileau}, \citenamefont {Christensen}, \citenamefont {Meyer},\ and\
  \citenamefont {Cornish}}]{Boileau:2020rpg}%
  \BibitemOpen
  \bibfield  {author} {\bibinfo {author} {\bibfnamefont {G.}~\bibnamefont
  {Boileau}}, \bibinfo {author} {\bibfnamefont {N.}~\bibnamefont
  {Christensen}}, \bibinfo {author} {\bibfnamefont {R.}~\bibnamefont {Meyer}},\
  and\ \bibinfo {author} {\bibfnamefont {N.~J.}\ \bibnamefont {Cornish}},\
  }\bibfield  {title} {\bibinfo {title} {{Spectral separation of the stochastic
  gravitational-wave background for LISA: Observing both cosmological and
  astrophysical backgrounds}},\ }\href
  {https://doi.org/10.1103/PhysRevD.103.103529} {\bibfield  {journal} {\bibinfo
   {journal} {Phys. Rev. D}\ }\textbf {\bibinfo {volume} {103}},\ \bibinfo
  {pages} {103529} (\bibinfo {year} {2021})},\ \Eprint
  {https://arxiv.org/abs/2011.05055} {arXiv:2011.05055 [gr-qc]} \BibitemShut
  {NoStop}%
\bibitem [{\citenamefont {Smith}\ and\ \citenamefont
  {Caldwell}(2019)}]{Smith:2019wny}%
  \BibitemOpen
  \bibfield  {author} {\bibinfo {author} {\bibfnamefont {T.~L.}\ \bibnamefont
  {Smith}}\ and\ \bibinfo {author} {\bibfnamefont {R.}~\bibnamefont
  {Caldwell}},\ }\bibfield  {title} {\bibinfo {title} {{LISA for Cosmologists:
  Calculating the Signal-to-Noise Ratio for Stochastic and Deterministic
  Sources}},\ }\href {https://doi.org/10.1103/PhysRevD.100.104055} {\bibfield
  {journal} {\bibinfo  {journal} {Phys. Rev. D}\ }\textbf {\bibinfo {volume}
  {100}},\ \bibinfo {pages} {104055} (\bibinfo {year} {2019})},\ \Eprint
  {https://arxiv.org/abs/1908.00546} {arXiv:1908.00546 [astro-ph.CO]}
  \BibitemShut {NoStop}%
\bibitem [{\citenamefont {Saffer}\ and\ \citenamefont
  {Yagi}(2020)}]{Saffer:2020xsw}%
  \BibitemOpen
  \bibfield  {author} {\bibinfo {author} {\bibfnamefont {A.}~\bibnamefont
  {Saffer}}\ and\ \bibinfo {author} {\bibfnamefont {K.}~\bibnamefont {Yagi}},\
  }\bibfield  {title} {\bibinfo {title} {{Parameter Estimation for Tests of
  General Relativity with the Astrophysical Stochastic Gravitational Wave
  Background}},\ }\href {https://doi.org/10.1103/PhysRevD.102.024001}
  {\bibfield  {journal} {\bibinfo  {journal} {Phys. Rev. D}\ }\textbf {\bibinfo
  {volume} {102}},\ \bibinfo {pages} {024001} (\bibinfo {year} {2020})},\
  \Eprint {https://arxiv.org/abs/2003.11128} {arXiv:2003.11128 [gr-qc]}
  \BibitemShut {NoStop}%
\bibitem [{\citenamefont {Harris}\ \emph {et~al.}(2020)\citenamefont {Harris},
  \citenamefont {Millman}, \citenamefont {van~der Walt}, \citenamefont
  {Gommers}, \citenamefont {Virtanen}, \citenamefont {Cournapeau},
  \citenamefont {Wieser}, \citenamefont {Taylor}, \citenamefont {Berg},
  \citenamefont {Smith}, \citenamefont {Kern}, \citenamefont {Picus},
  \citenamefont {Hoyer}, \citenamefont {van Kerkwijk}, \citenamefont {Brett},
  \citenamefont {Haldane}, \citenamefont {del R{\'{i}}o}, \citenamefont
  {Wiebe}, \citenamefont {Peterson}, \citenamefont {G{\'{e}}rard-Marchant},
  \citenamefont {Sheppard}, \citenamefont {Reddy}, \citenamefont {Weckesser},
  \citenamefont {Abbasi}, \citenamefont {Gohlke},\ and\ \citenamefont
  {Oliphant}}]{harris2020array}%
  \BibitemOpen
  \bibfield  {author} {\bibinfo {author} {\bibfnamefont {C.~R.}\ \bibnamefont
  {Harris}}, \bibinfo {author} {\bibfnamefont {K.~J.}\ \bibnamefont {Millman}},
  \bibinfo {author} {\bibfnamefont {S.~J.}\ \bibnamefont {van~der Walt}},
  \bibinfo {author} {\bibfnamefont {R.}~\bibnamefont {Gommers}}, \bibinfo
  {author} {\bibfnamefont {P.}~\bibnamefont {Virtanen}}, \bibinfo {author}
  {\bibfnamefont {D.}~\bibnamefont {Cournapeau}}, \bibinfo {author}
  {\bibfnamefont {E.}~\bibnamefont {Wieser}}, \bibinfo {author} {\bibfnamefont
  {J.}~\bibnamefont {Taylor}}, \bibinfo {author} {\bibfnamefont
  {S.}~\bibnamefont {Berg}}, \bibinfo {author} {\bibfnamefont {N.~J.}\
  \bibnamefont {Smith}}, \bibinfo {author} {\bibfnamefont {R.}~\bibnamefont
  {Kern}}, \bibinfo {author} {\bibfnamefont {M.}~\bibnamefont {Picus}},
  \bibinfo {author} {\bibfnamefont {S.}~\bibnamefont {Hoyer}}, \bibinfo
  {author} {\bibfnamefont {M.~H.}\ \bibnamefont {van Kerkwijk}}, \bibinfo
  {author} {\bibfnamefont {M.}~\bibnamefont {Brett}}, \bibinfo {author}
  {\bibfnamefont {A.}~\bibnamefont {Haldane}}, \bibinfo {author} {\bibfnamefont
  {J.~F.}\ \bibnamefont {del R{\'{i}}o}}, \bibinfo {author} {\bibfnamefont
  {M.}~\bibnamefont {Wiebe}}, \bibinfo {author} {\bibfnamefont
  {P.}~\bibnamefont {Peterson}}, \bibinfo {author} {\bibfnamefont
  {P.}~\bibnamefont {G{\'{e}}rard-Marchant}}, \bibinfo {author} {\bibfnamefont
  {K.}~\bibnamefont {Sheppard}}, \bibinfo {author} {\bibfnamefont
  {T.}~\bibnamefont {Reddy}}, \bibinfo {author} {\bibfnamefont
  {W.}~\bibnamefont {Weckesser}}, \bibinfo {author} {\bibfnamefont
  {H.}~\bibnamefont {Abbasi}}, \bibinfo {author} {\bibfnamefont
  {C.}~\bibnamefont {Gohlke}},\ and\ \bibinfo {author} {\bibfnamefont {T.~E.}\
  \bibnamefont {Oliphant}},\ }\bibfield  {title} {\bibinfo {title} {Array
  programming with {NumPy}},\ }\href
  {https://doi.org/10.1038/s41586-020-2649-2} {\bibfield  {journal} {\bibinfo
  {journal} {Nature}\ }\textbf {\bibinfo {volume} {585}},\ \bibinfo {pages}
  {357} (\bibinfo {year} {2020})}\BibitemShut {NoStop}%
\bibitem [{\citenamefont {Virtanen}\ \emph {et~al.}(2020)\citenamefont
  {Virtanen}, \citenamefont {Gommers}, \citenamefont {Oliphant}, \citenamefont
  {Haberland}, \citenamefont {Reddy}, \citenamefont {Cournapeau}, \citenamefont
  {Burovski}, \citenamefont {Peterson}, \citenamefont {Weckesser},
  \citenamefont {Bright}, \citenamefont {{van der Walt}}, \citenamefont
  {Brett}, \citenamefont {Wilson}, \citenamefont {Millman}, \citenamefont
  {Mayorov}, \citenamefont {Nelson}, \citenamefont {Jones}, \citenamefont
  {Kern}, \citenamefont {Larson}, \citenamefont {Carey}, \citenamefont {Polat},
  \citenamefont {Feng}, \citenamefont {Moore}, \citenamefont {{VanderPlas}},
  \citenamefont {Laxalde}, \citenamefont {Perktold}, \citenamefont {Cimrman},
  \citenamefont {Henriksen}, \citenamefont {Quintero}, \citenamefont {Harris},
  \citenamefont {Archibald}, \citenamefont {Ribeiro}, \citenamefont
  {Pedregosa}, \citenamefont {{van Mulbregt}},\ and\ \citenamefont {{SciPy 1.0
  Contributors}}}]{2020SciPy-NMeth}%
  \BibitemOpen
  \bibfield  {author} {\bibinfo {author} {\bibfnamefont {P.}~\bibnamefont
  {Virtanen}}, \bibinfo {author} {\bibfnamefont {R.}~\bibnamefont {Gommers}},
  \bibinfo {author} {\bibfnamefont {T.~E.}\ \bibnamefont {Oliphant}}, \bibinfo
  {author} {\bibfnamefont {M.}~\bibnamefont {Haberland}}, \bibinfo {author}
  {\bibfnamefont {T.}~\bibnamefont {Reddy}}, \bibinfo {author} {\bibfnamefont
  {D.}~\bibnamefont {Cournapeau}}, \bibinfo {author} {\bibfnamefont
  {E.}~\bibnamefont {Burovski}}, \bibinfo {author} {\bibfnamefont
  {P.}~\bibnamefont {Peterson}}, \bibinfo {author} {\bibfnamefont
  {W.}~\bibnamefont {Weckesser}}, \bibinfo {author} {\bibfnamefont
  {J.}~\bibnamefont {Bright}}, \bibinfo {author} {\bibfnamefont {S.~J.}\
  \bibnamefont {{van der Walt}}}, \bibinfo {author} {\bibfnamefont
  {M.}~\bibnamefont {Brett}}, \bibinfo {author} {\bibfnamefont
  {J.}~\bibnamefont {Wilson}}, \bibinfo {author} {\bibfnamefont {K.~J.}\
  \bibnamefont {Millman}}, \bibinfo {author} {\bibfnamefont {N.}~\bibnamefont
  {Mayorov}}, \bibinfo {author} {\bibfnamefont {A.~R.~J.}\ \bibnamefont
  {Nelson}}, \bibinfo {author} {\bibfnamefont {E.}~\bibnamefont {Jones}},
  \bibinfo {author} {\bibfnamefont {R.}~\bibnamefont {Kern}}, \bibinfo {author}
  {\bibfnamefont {E.}~\bibnamefont {Larson}}, \bibinfo {author} {\bibfnamefont
  {C.~J.}\ \bibnamefont {Carey}}, \bibinfo {author} {\bibfnamefont
  {{\.I}.}~\bibnamefont {Polat}}, \bibinfo {author} {\bibfnamefont
  {Y.}~\bibnamefont {Feng}}, \bibinfo {author} {\bibfnamefont {E.~W.}\
  \bibnamefont {Moore}}, \bibinfo {author} {\bibfnamefont {J.}~\bibnamefont
  {{VanderPlas}}}, \bibinfo {author} {\bibfnamefont {D.}~\bibnamefont
  {Laxalde}}, \bibinfo {author} {\bibfnamefont {J.}~\bibnamefont {Perktold}},
  \bibinfo {author} {\bibfnamefont {R.}~\bibnamefont {Cimrman}}, \bibinfo
  {author} {\bibfnamefont {I.}~\bibnamefont {Henriksen}}, \bibinfo {author}
  {\bibfnamefont {E.~A.}\ \bibnamefont {Quintero}}, \bibinfo {author}
  {\bibfnamefont {C.~R.}\ \bibnamefont {Harris}}, \bibinfo {author}
  {\bibfnamefont {A.~M.}\ \bibnamefont {Archibald}}, \bibinfo {author}
  {\bibfnamefont {A.~H.}\ \bibnamefont {Ribeiro}}, \bibinfo {author}
  {\bibfnamefont {F.}~\bibnamefont {Pedregosa}}, \bibinfo {author}
  {\bibfnamefont {P.}~\bibnamefont {{van Mulbregt}}},\ and\ \bibinfo {author}
  {\bibnamefont {{SciPy 1.0 Contributors}}},\ }\bibfield  {title} {\bibinfo
  {title} {{{SciPy} 1.0: Fundamental Algorithms for Scientific Computing in
  Python}},\ }\href {https://doi.org/10.1038/s41592-019-0686-2} {\bibfield
  {journal} {\bibinfo  {journal} {Nature Methods}\ }\textbf {\bibinfo {volume}
  {17}},\ \bibinfo {pages} {261} (\bibinfo {year} {2020})}\BibitemShut
  {NoStop}%
\bibitem [{\citenamefont {Hunter}(2007)}]{Hunter:2007ouj}%
  \BibitemOpen
  \bibfield  {author} {\bibinfo {author} {\bibfnamefont {J.~D.}\ \bibnamefont
  {Hunter}},\ }\bibfield  {title} {\bibinfo {title} {{Matplotlib: A 2D Graphics
  Environment}},\ }\href {https://doi.org/10.1109/MCSE.2007.55} {\bibfield
  {journal} {\bibinfo  {journal} {Comput. Sci. Eng.}\ }\textbf {\bibinfo
  {volume} {9}},\ \bibinfo {pages} {90} (\bibinfo {year} {2007})}\BibitemShut
  {NoStop}%
\bibitem [{\citenamefont {Lewis}(2019)}]{Lewis:2019xzd}%
  \BibitemOpen
  \bibfield  {author} {\bibinfo {author} {\bibfnamefont {A.}~\bibnamefont
  {Lewis}},\ }\bibfield  {title} {\bibinfo {title} {{GetDist: a Python package
  for analysing Monte Carlo samples}},\ }\href@noop {} {\  (\bibinfo {year}
  {2019})},\ \Eprint {https://arxiv.org/abs/1910.13970} {arXiv:1910.13970
  [astro-ph.IM]} \BibitemShut {NoStop}%
\bibitem [{\citenamefont {Christensen}(1992)}]{Christensen:1992wi}%
  \BibitemOpen
  \bibfield  {author} {\bibinfo {author} {\bibfnamefont {N.}~\bibnamefont
  {Christensen}},\ }\bibfield  {title} {\bibinfo {title} {{Measuring the
  stochastic gravitational radiation background with laser interferometric
  antennas}},\ }\href {https://doi.org/10.1103/PhysRevD.46.5250} {\bibfield
  {journal} {\bibinfo  {journal} {Phys. Rev. D}\ }\textbf {\bibinfo {volume}
  {46}},\ \bibinfo {pages} {5250} (\bibinfo {year} {1992})}\BibitemShut
  {NoStop}%
\end{thebibliography}%

\end{document}